\documentclass[11pt,a4paper]{article}
\pdfoutput=1
\usepackage{amssymb} \usepackage{amsmath} \usepackage{graphicx}
\usepackage{epsfig,latexsym,color,xcolor}
\usepackage{ulem}
\usepackage{amstext}    
\usepackage{array} 
\begin{document}
\def\Giulia{\bf\color{red}}
\def\bef{\begin{figure}}
\def\eef{\end{figure}}
\newcommand{\ans}{ansatz }
\newcommand{\be}[1]{\begin{equation}\label{#1}}
\newcommand{\beq}{\begin{equation}}
\newcommand{\ee}{\end{equation}}
\newcommand{\beqn}[1]{\begin{eqnarray}\label{#1}}
\newcommand{\eeqn}{\end{eqnarray}}
\newcommand{\bd}{\begin{displaymath}}
\newcommand{\ed}{\end{displaymath}}
\newcommand{\mat}[4]{\left(\begin{array}{cc}{#1}&{#2}\\{#3}&{#4}
\end{array}\right)}
\newcommand{\matr}[9]{\left(\begin{array}{ccc}{#1}&{#2}&{#3}\\
{#4}&{#5}&{#6}\\{#7}&{#8}&{#9}\end{array}\right)}
\newcommand{\matrr}[6]{\left(\begin{array}{cc}{#1}&{#2}\\
{#3}&{#4}\\{#5}&{#6}\end{array}\right)}
\newcommand{\cvb}[3]{#1^{#2}_{#3}}
\def\lsim{\raise0.3ex\hbox{$\;<$\kern-0.75em\raise-1.1ex
e\hbox{$\sim\;$}}}
\def\gsim{\raise0.3ex\hbox{$\;>$\kern-0.75em\raise-1.1ex
\hbox{$\sim\;$}}}
\def\abs#1{\left| #1\right|}
\def\simlt{\mathrel{\lower2.5pt\vbox{\lineskip=0pt\baselineskip=0pt
           \hbox{$<$}\hbox{$\sim$}}}}
\def\simgt{\mathrel{\lower2.5pt\vbox{\lineskip=0pt\baselineskip=0pt
           \hbox{$>$}\hbox{$\sim$}}}}
\def\unity{{\hbox{1\kern-.8mm l}}}
\newcommand{\eps}{\varepsilon}
\def\ep{\epsilon}
\def\ga{\gamma}
\def\Ga{\Gamma}
\def\om{\omega}
\def\omp{{\omega^\prime}}
\def\Om{\Omega}
\def\la{\lambda}
\def\La{\Lambda}
\def\al{\alpha}
\def\beq{\begin{equation}}
\def\eeq{\end{equation}}
\newcommand{\ov}{\overline}
\renewcommand{\to}{\rightarrow}
\renewcommand{\vec}[1]{\mathbf{#1}}
\newcommand{\vect}[1]{\mbox{\boldmath$#1$}}
\def\tm{{\widetilde{m}}}
\def\mcirc{{\stackrel{o}{m}}}
\newcommand{\Dm}{\Delta m}
\newcommand{\dm}{\varepsilon}
\newcommand{\tanb}{\tan\beta}
\newcommand{\nbar}{\tilde{n}}
\newcommand\PM[1]{\begin{pmatrix}#1\end{pmatrix}}
\newcommand{\up}{\uparrow}
\newcommand{\down}{\downarrow}
\newcommand{\refs}[2]{eqs.~(\ref{#1})-(\ref{#2})}
\def\omE{\omega_{\rm Ter}}
\newcommand{\eqn}[1]{eq.~(\ref{#1})}
%

\newcommand{\DSUSY}{{SUSY \hspace{-9.4pt} \slash}\;}
\newcommand{\DCP}{{CP \hspace{-7.4pt} \slash}\;}
\newcommand{\mc}{\mathcal}
\newcommand{\gr}{\mathbf}
\renewcommand{\to}{\rightarrow}
\newcommand{\gtc}{\mathfrak}
\newcommand{\wh}{\widehat}
\newcommand{\br}{\langle}
\newcommand{\kt}{\rangle}


\def\lsim{\mathrel{\mathop  {\hbox{\lower0.5ex\hbox{$\sim$}
\kern-0.8em\lower-0.7ex\hbox{$<$}}}}}
\def\gsim{\mathrel{\mathop  {\hbox{\lower0.5ex\hbox{$\sim$}
\kern-0.8em\lower-0.7ex\hbox{$>$}}}}}

\def\nn{\\  \nonumber}
\def\de{\partial}
\def\brf{{\mathbf f}}
\def\bbf{\bar{\bf f}}
\def\bF{{\bf F}}
\def\bbF{\bar{\bf F}}
\def\bA{{\mathbf A}}
\def\bB{{\mathbf B}}
\def\bG{{\mathbf G}}
\def\bI{{\mathbf I}}
\def\bM{{\mathbf M}}
\def\bY{{\mathbf Y}}
\def\bX{{\mathbf X}}
\def\bS{{\mathbf S}}
\def\bb{{\mathbf b}}
\def\bh{{\mathbf h}}
\def\bg{{\mathbf g}}
\def\bla{{\mathbf \la}}
\def\bmu{\mathbf m }
\def\by{{\mathbf y}}
\def\bmu{\mbox{\boldmath $\mu$} }
\def\bsig{\mbox{\boldmath $\sigma$} }
\def\bunity{{\mathbf 1}}
\def\cA{{\cal A}}
\def\cB{{\cal B}}
\def\cC{{\cal C}}
\def\cD{{\cal D}}
\def\cF{{\cal F}}
\def\cG{{\cal G}}
\def\cH{{\cal H}}
\def\cI{{\cal I}}
\def\cL{{\cal L}}
\def\cN{{\cal N}}
\def\cM{{\cal M}}
\def\cO{{\cal O}}
\def\cR{{\cal R}}
\def\cS{{\cal S}}
\def\cT{{\cal T}}
\def\eV{{\rm eV}}
%




\large
 \begin{center}
 {\Large \bf Exotic see-saw mechanism for neutrinos and leptogenesis \\
 in a Pati-Salam model}
 \end{center}

 \vspace{0.1cm}

 \vspace{0.1cm}
 \begin{center}
{\large Andrea Addazi}\footnote{E-mail: \,  andrea.addazi@infn.lngs.it} \\
{\it \it Dipartimento di Fisica,
 Universit\`a di L'Aquila, 67010 Coppito, AQ \\
LNGS, Laboratori Nazionali del Gran Sasso, 67010 Assergi AQ, Italy}
\end{center}

  \begin{center}
{\large Massimo Bianchi}\footnote{E-mail: \, massimo.bianchi@roma2.infn.it}
\\
{\it Dipartimento di Fisica, Universit\`a di Roma Tor Vergata, \\
I.N.F.N. Sezione di Roma Tor Vergata, \\
Via della Ricerca Scientifica, 1 00133 Roma, Italy}
\end{center}

\begin{center}
{\large Giulia Ricciardi}\footnote{E-mail: \, giulia.ricciardi@na.infn.it}
\\
{\it 
Dipartimento di Fisica E. Pancini, Universit\`a di Napoli Federico II, \\ 
I.N.F.N. Sezione di Napoli, \\  
Complesso Universitario di Monte Sant'Angelo, Via Cintia, 80126 Napoli, Italy}

\end{center}

\vspace{1cm}
\begin{abstract}
\large

We discuss non-perturbative corrections 
to the neutrino sector, in the context 
of a D-brane Pati-Salam-like model, that can be obtained as a  
simple alternative to $SO(10)$ GUT's in theories with open and unoriented strings.
In such D-brane models, exotic stringy instantons
can correct the right-handed neutrino mass matrix in a calculable way, thus affecting mass hierarchies and modifying the see-saw mechanism to what we name {\it exotic see-saw}.
 For a wide range of parameters, a compact spectrum 
of right-handed neutrino masses
can occur that gives rise to 
a predictive scenario for low energy observables.
This model also provides a 
viable mechanism for Baryon Asymmetry in the Universe (BAU) through leptogenesis. 
Finally, a Majorana mass for the neutron
is naturally predicted in the model, leading to
potentially testable neutron-antineutron oscillations. 
Combined measurements in neutrino and neutron-antineutron sectors 
could provide precious informations on physics at the quantum gravity scale.  

\end{abstract}

\baselineskip = 20pt

\section{Introduction}
 
In \cite{Majorana}, Majorana proposed the existence of 
extra mass terms of the form $m \psi \psi+h.c$, 
in which $\psi$ is a neutral  fermion, such 
as a neutrino or a neutron. 
Majorana's  proposal  has never seemed to be so up-to-date and intriguing
as today. 
In fact, from several measures 
of atmospheric, solar, accelerator and reactor neutrinos , 
neutrino oscillations have been fully confirmed.
These observations represent evidence that neutrinos  are massive. 
Majorana's proposal goes even beyond 
the  mass issues: a Majorana mass term for neutrinos or for the neutron leads 
to violation of Lepton (L) and Baryon (B) numbers 
as $\Delta L=2$ and $\Delta B=2$, respectively.
The Standard Model (SM) does not offer an adequate explanation of the observed Matter-Antimatter asymmetry in our Universe,
{\it i.~e.}  the SM does not generate the necessary Lepton and/or Baryon number asymmetries 
in the primordial Universe.
The possibility of a  Majorana mass term
 for neutrino or neutron  
 can disclose new paths towards the origin of the observed asymmetry and its possible dynamical generation, through a viable mechanism for baryogenesis.

 See-saw Type I mechanism is considered one of the most elegant ways to explain
the observed smallness of neutrino masses \cite{ss1, GellMann:1980vs,  ss2,ss3,ss4}. In see-saw Type I, right-handed (RH) neutrinos with masses much higher than the
electroweak (EW) scale are  required. 
Remarkably, this mechanism offers
a simple and natural solution for leptogenesis, a model of baryogenesis where
the lightest RH neutrino can decay into lighter particles \cite{FY}.
In the primordial universe, near the EW phase transition, leptons, quarks and Higgs also interact via $B+L$ violating non-perturbative interactions,
generated by {\it sphalerons}, leading to an effective conversion of part of the initial lepton number asymmetry into a  baryonic one \cite{Sph2}. 
 Moreover,  the complex Yukawa couplings of the RH
neutrinos  can provide new sources of CP violation.
All Sakharov's conditions to dynamically generate baryon asymmetry \cite{Sakharov}  are satisfied: 
1) out of thermal equilibrium condition; 2)  
CP violations; 3) baryon number violation.
The sphaleron-mediated effective interactions were calculated for the first time by t'Hooft \cite{Sph1}. These effects  are strongly suppressed 
in our present cosmological epoch but, in the primordial thermal bath, they are expected to be unsuppressed, leading to non-negligible corrections to the chemical potentials.  

The see-saw mechanism can be naturally embedded in a Pati-Salam (PS)
model $SU(3)_{c}{\times}  SU(2)_{L}{\times}  SU(2)_{R}{\times}  U(1)_{B-L}$
or $SU(4)_{c}{\times}  Sp(2)_{L}{\times}  Sp(2)_{R}$ \cite{PS}.
As suggested in \cite{ss4} Majorana masses for neutrinos  can be elegantly connected to a spontaneous symmetry breaking of parity and to leptogenesis.  
In fact the RH masses are related to Left-Right scale and $U(1)_{B-L}\subset SU(4)_{c}$ spontaneous symmetry breaking scale. 
On the other hand, a RH neutrino  mass scale of order $M_{R}\sim 10^{9\div 13}\,\rm GeV$ is necessary for consistent leptogenesis \cite{DI}.

As a natural step beyond a PS-model,
$SO(10)$ GUT could unify the SM with $U(1)_{B-L}$ via an intermediate
$SU(4)_{c}{\times}  SU(2)_{L}{\times}  SU(2)_R$ PS-like gauge group \footnote{Recent discussions about 
 $SO(10)$ GUT 
 can be found in \cite{SO10a,SO10b,SO10c,SO10d,SO10e}}.  
 However, let us recall that the $SO(10)$ GUT scenario presents some 
challenging theoretical problems, that  are generally  solved  at the cost of
some complications of the initial GUT model.
Problems   such as proton destabilization and 
the imperfect unification of coupling constants are generally alleviated
in SUSY $SO(10)$ GUT. 
With or without SUSY, the most serious hierarchy problem for $SO(10)$ and other GUTs is the {\it doublet-triplet splitting}. The standard Higgs doublet 
is contained in ${10}_{H}$ (or ${5}_{H} + {5}^*_{H}$ in $SU(5)$), leading to dangerous scale-mixing  diagrams between standard doublets and heavier Higgs triplets inside ${10}_{H}$. In other words, a stabilization of the ordinary doublet at much smaller scales than $M_{GUT}\simeq 10^{15\div 16}\,\rm GeV$ is highly unnatural, {\it i.~e.} it reintroduces another Higgs hierarchy problem  
even if one assumes $1\, \rm TeV$ SUSY breaking scale\footnote{The doublet-triplet problem can be solved in different ways
in GUT models. 
The most popular solution is the missing partner or vacuum-expectation-value mechanism for SU(5) \cite{dt1}. In $SO(10)$, an implementation of this mechanism 
was shown in \cite{BMt}. 
As an alternative, we mention 
 pseudo-Nambu-Goldstone boson mechanism for
SU(6) \cite{NG, NG1, NG2}. Finally, in string theory (and orbifold GUTs),  orbifold projection can remove Higgs triplets \cite{stringGUT, stringGUT1,stringGUT2,stringGUT3,stringGUT4}. }.

In $SO(10)$, the quark-lepton symmetry makes 
the reconciliation of leptogenesis and see-saw mechanism more problematic.
In fact,
assuming the spontaneous symmetry breaking scale of $SU(4)_{c}{\times}  SU(2)_{L}{\times}  SU(2)_{R} \rightarrow SU(3)_{c}{\times}  SU(2)_{L}{\times}  U(1)_{Y}$ 
around $\Lambda_{R}\simeq 10^{11}\, \rm GeV$, 
the lightest RH eigenstate $N_{1}$, which is generally the main responsible for generating a lepton asymmetry,  acquires a mass 
$M_{R_{1}} \ll 10^{9}\,\rm GeV$.
Unfortunately, this value is well below the Davidson-Ibarra (DI) bound \cite{DI}
($M_{DI}\gtrsim 10^{9}\, \rm GeV$),
 guaranteeing a sufficient production of lepton asymmetry from RH neutrino decays. 
There are basically  three ways out of this difficulty.
One possibility is to consider leptogenesis where crucial contributions arise via the decays of heavier RH neutrinos , with masses   above the DI limit  \cite{11,26,27,28,29,30,31}.
 Alternatively, one can assume a highly compact spectrum \cite{7,Buccella:2012kc}.
Finally,  in a situation in which one pair of RH neutrinos  
 is highly degenerate, the DI bound can be avoided  through 
 a resonant enhancement of CP asymmetries \cite{32a,32b}. 
Let us observe that the latter two scenarios
are not easily incorporated in $SO(10)$\footnote{For recent literature 
discussing these aspects, see
\cite{m1,m2,m3}.}.

Lastly, it is undoubtable that $SO(10)$
cannot provide a way to unify gravity with the other interactions.
Indeed, $SO(10)$ scenarios are not the only possible completion 
of PS-like models. 
In IIA and IIB superstring theory, a natural way to construct 
a PS-like model can be achieved through a system of intersecting D-branes 
stacks wrapping some sub-manifold (`cycles') in a Calabi-Yau (CY) compactifications with open strings ending 
on them. In this class of models, a different kind of {\it geometric unification} can be achieved, including gravity -- even if string theory were 
incomplete, even if quantum gravity were only understood partially 
\footnote{As in GUTs, also in these models we can find some 
difficult theoretical problems: i) the identification of the precise CY singularity for the D-brane construction, ii) the quantitative stabilization of geometric moduli for the particular realistic particle physics model considered. These problems are expected to be solved by 
including fluxes and the effects of stringy instantons.
For the moment, awaiting for a more precise quantitative UV completion (global embedding)
of our model, we can neglect these problematics.
Our attitude is to consider effective string-inspired models,
locally free from anomalies and tadpoles and interesting for phenomenology of particle physics and cosmology. On the other hand, attempts  
to solve the problems mentioned above are the main topics of 
an intense investigation. For example, 
see \cite{Quevedo1,Quevedo2,Quevedo3} for recent discussions. }.
Recently, a simple D-branes PS-like model was suggested in \cite{Addazi:2015hka}.
In \cite{Addazi:2015hka}, we have noticed that a Higgs sector composed of 
$\Delta(10,1,1),\Delta^{c}(10^{*},1,1)$, $\phi_{LL}(1,3,1),\phi_{RR}(1,1,3)$ and $h_{LR}(1,2,2)$, the latter containing SM Higgses,
can reproduce the right pattern of fermion masses. 
However, the above Higgses cannot break $SU(4){\times}  SU(2)_{R}$ down 
to $SU(3){\times}  U(1)_{Y}$ in the desired way. 
This spontaneous symmetry breaking can be obtained 
through Higgs superfields $\bar{H}(\bar{4},1,2)$ and $H(4,1,2)$.
In $SO(10)$, they are usually contained in $16_{H},\bar{16}_{H}$.
$\bar{H}$ has the same representation $F_{R}$ of the standard fermions and their super-partners, while $H$ is in the conjugate one. 
They can be decomposed in components as 
\be{H}
\bar{H}(\bar{4},1,2)=(u_{R}^{c},d_{R}^{c},e_{R}^{c},\nu_{R}^{c})
\ee
\be{Hbar}
H(4,1,2)=(\bar{u}_{R}^{c},\bar{d}_{R}^{c},\bar{e}_{R}^{c},\bar{\nu}_{R}^{c})
\ee
The vacuum expectation values  (VEVs) along the ``sneutrino'' components
\be{vevsHandHbar}
\langle \bar{H}\rangle=\langle \nu_{R}^{c} \rangle
\qquad , \qquad
\langle H \rangle=\langle \bar{\nu}_{R}^{c} \rangle 
\ee
break $SU(4){\times}  SU(2)_{R}$ to $SU(3){\times}  U(1)_{Y}$.
VEVs (\ref{vevsHandHbar}) 
have to be higher than $\langle \Delta^{c} \rangle, \langle \phi_{RR} \rangle$ 
in order to guarantee the right symmetry breaking pattern\footnote{For this reason, a TeV-ish Left-Right symmetry breaking 
is not favored by our precise model. Comments on phenomenological aspects 
made in  \cite{Addazi:2015hka} can be valid in quivers 
inspired by the present one but with extra nodes.}.
In this model
a Majorana mass for the neutron and extra terms in the RH neutrino  mass matrix are generated by 
Euclidean D2-branes (or E2-branes),
wrapping a different 3-cycle with respect to the 
ordinary D6-branes. Such E2's are called
exotic instantons. They are a different kind of instantons not 
present in gauge theories. The effect of E2s 
are calculable and controllable in models like our one.
Unlike `gauge' instantons, `exotic' instantons do not admit an ADHM construction. See \cite{Bianchi:2009ij, Bianchi:2012ud} for useful reviews of these aspects \footnote{See \cite{Franco:2015kfa} for a recent paper on
D-brane instantons in chiral quiver theories.}. 
The main new peculiar feature of exotic instantons is that they can violate vector-like symmetries 
 like baryon and lepton numbers! 
 B/L-violations by exotic E2-instantons are not necessarily suppressed:
  suppression factors depend on the particular size of the
  3-cycles wrapped in the CY compactification by exotic E2-instantons. 
  A dynamical violation of a symmetry is something ``smarter" than an explicit one: all possible dangerous operators 
 are not generated by exotic instantons, only few interesting operators 
 can be generated. For instance, an effective operator 
 $(u^{c}d^{c}d^{c})^{2}/\Lambda_{n\bar{n}}^{5}$ is generated in our model,
 without proton destabilization: a residual discrete symmetry is preserved by exotic instantons, avoiding $\Delta B=1$ processes  but allowing 
$n{-}\bar{n}$ ($\Delta B=2$) transitions \cite{MM}. In particular, such transitions 
are mediated by 
  three color scalar sextets present in our model.
   E2-instantons generate an effective superpotential term $\mathcal{W}_{E2}=\Delta_{u^{c}u^{c}}^{(6)} \Delta_{d^{c}d^{c}}^{(6)} \Delta_{d^{c}d^{c}}^{(6)} \langle S^{(1)}\rangle/\mathcal{M}_{E}$, where 
  $\Delta_{6}=(6,1)_{+2/3}$ and $S=(1,1)_{-2}$ are contained 
  in $(10,1,1)$ of $SU(4)_{c}{\times}  Sp(2)_{L}{\times}  Sp(2)_{R}$. When $S$ takes an expectation value, spontaneously breaking 
  $U(1)_{B-L}$, an effective trilinear interaction for $\Delta^{(6)}$s is generated at low energies 
 of order  $\mathcal{M}_{E}\sim M_{S}$, 
  where $M_{S}$ is the string scale.
$n{-}\bar{n}$ transition can be obtained from $\mathcal{W}_{E2}$
  and renormalizable operators, present in our model and coded in a quiver,
 $\Delta^{(6)}_{u^{c}u^{c}}u^{c}u^{c}$
 and $\Delta^{(6)}_{d^{c}d^{d}}d^{c}d^{c}$, 
 with $\Lambda_{n\bar{n}}^{5}\simeq \mathcal{M}_{E}M_{\Delta_{u^{c}u^{c}}}^{2}M_{\Delta_{d^{c}d^{c}}}^{2}M_{SUSY}/v_{B-L}$
 where $M_{SUSY}$ is the SUSY breaking scale,
 $v_{B-L}$ the $U(1)_{B-L}$ breaking VEV.
Its scale can be as low as $\Lambda\simeq 1000\, \rm TeV$,
corresponding to $n{-}\bar{n}$ transitions in vacuum 
(no magnetic-fields, outside nuclei)
with $\tau_{n\bar{n}}\simeq 100\, \rm yr$,
{\it i.e} $10^{-33}\tau_{p-decay}$ \cite{PDG}.
The next generation of experiments 
promises to test exactly this scale, 
enhancing the current best limits for $\tau_{n\bar{n}}$ \cite{Baldo}
by two orders of magnitude \cite{Modern1,Modern2}.
 In string theory, $M_{S}$ needs not be necessarily close to the Planck scale,  
it can easily stay at a lower scale. Similarly the SUSY breaking scale 
is not necessarily at the TeV scale - since we are only interested in 
SUSY as a symmetry for superstring theory, we will consider it to be around the String scale \footnote{An alternative mechanism for Baryon Asymmetry of the Universe (BAU) can be envisaged.
As proposed in \cite{MM,BM1,BM2,BM3}, a Post-Sphaleron Baryogenesis 
mediated by color scalar sextets could be a viable alternative 
to a Leptogenesis-Sphaleron mechanism. An intriguing possibility 
is to test this scenario in Neutron-Antineutron physics. 
Color scalar sextets are naturally embedded not only in 
$SO(10)$, but also in our model with intersecting D-branes, 
as extensively discussed in \cite{Addazi:2015hka}.}.
Direct limits on color sextet scalars can be obtained
from FCNCs as discussed in \cite{BFM1,BFM2},
usually stronger than LHC ones \cite{LHC,LHCexp}
\footnote{For other D-branes model generating a Majorana mass for the neutron and other intriguing signatures for phenomenology, 
in Ultra Cold Neutron Physics, Ultra High Energy Cosmic Rays, 
FCNCs and LHC, see \cite{Addazi:2014ila,Addazi:2015ata,Addazi:2015rwa,Addazi:2015eca,Addazi:2015fua,Addazi:2015oba,Addazi:2015goa}.}.
In the present paper,  we discuss 
quantitative predictions of our PS-like model
for low energy observables in neutrino physics,
as done in the literature for $SO(10)$ GUT's. 
We show that our model can be remarkably predictive for neutrino physics,
exposing a quark-lepton symmetry 
and a compact spectrum of RH neutrinos 
with masses above the DI bound 
for leptogenesis.
The compactness of the mass spectrum of RH neutrinos  
is related
to the geometrical proprieties of the relevant mixed disk amplitudes.
Our model provides a theoretical framework where a compact RH spectrum emerges naturally.
 In our phenomenological analysis, we will take into account a non vanishing value 
of the lepton mixing angle $\theta_{13}$, as measured
in \cite{13,14,15}, assuming the best fit value given in \cite{15}.
We will see how the compactness of the RH neutrino  mass spectrum 
leads to consistent solution with a non-zero Dirac phase $\delta\neq 0$, 
in the Pontecorvo-Maki-Nagakawa-Sakata (PMNS)
mixing matrix. The solutions obtained then fix 
the other unknown low energy parameters:
the PMNS CP violating phases $\delta,\alpha, \beta$  
(modulo signs) and the left-handed (LH) neutrino mass scale $M_{1}$. We also predict the RH neutrino masses.
The numerical approach follows  the path drawn in the context of $SO(10)$ GUT, where a compact RH spectrum represented 
a somewhat  arbitrary assumption \cite{7,Buccella:2012kc}.
The plan of the paper is as follows. In Sect.~\ref{PSlikeABDbrane} we review and amend 
a Pati-Salam-like model with gauge $U(4){\times}  Sp(2)_{L}{\times}  Sp(2)_{R}$ based on unoriented D-branes proposed in \cite{Addazi:2015hka}. 
In Sect.~\ref{LEPTO} we calculate relevant parameters for leptogenesis in a case where 
the right order of magnitude and sign of the BAU is recovered, a  non trivial result 
in view of the high level of predictability of the present model.

\section{Pati-Salam-like D-brane models}
\label{PSlikeABDbrane}

\begin{figure}[t]
\centerline{ \includegraphics [height=8cm,width=1.0 \columnwidth]{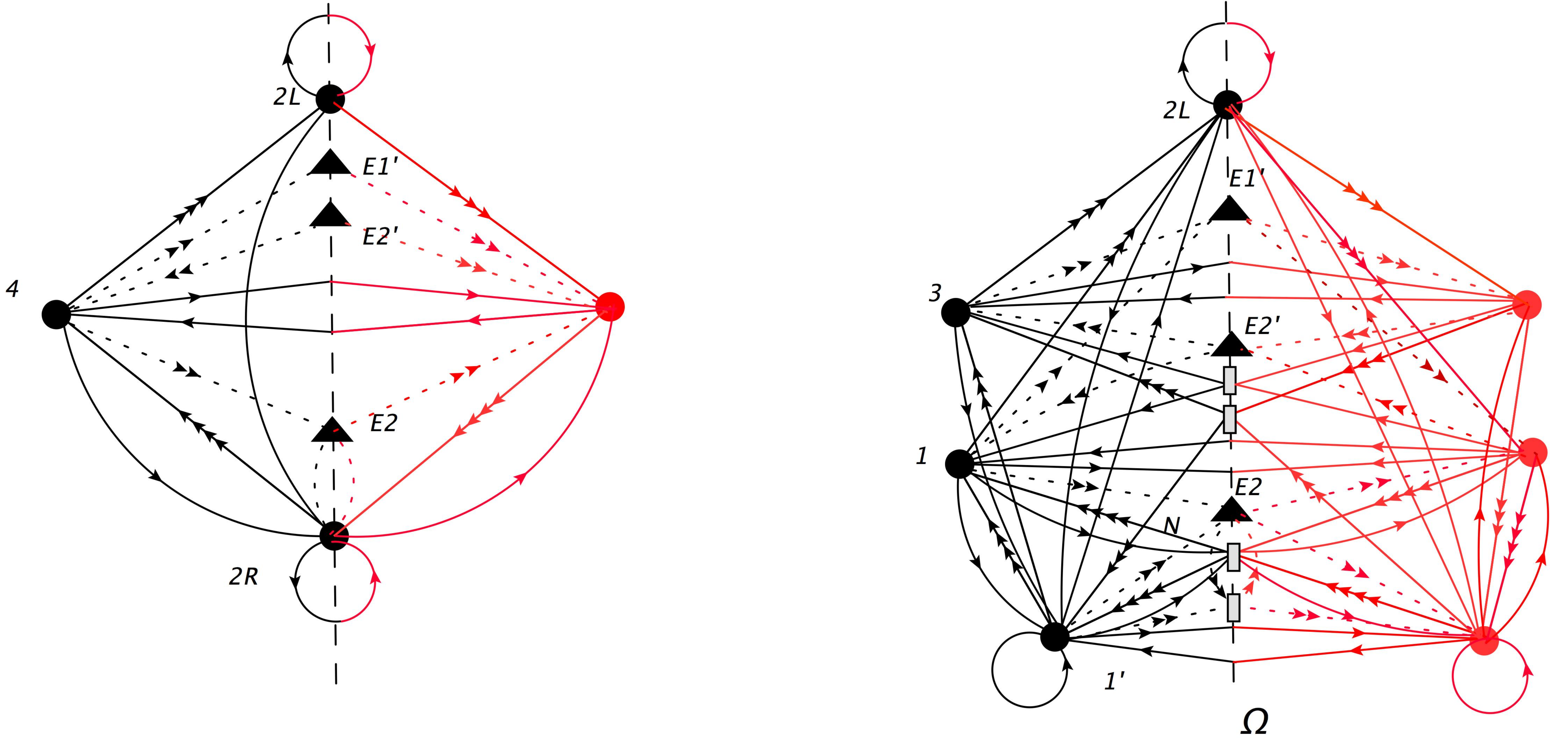}}
\vspace*{-1ex}
\caption{ On the left, the unoriented quiver for a Pati-Salam-like model $U(4){{\times} } Sp(2)_{L}{{\times} }Sp(2)_{R}$ is shown. Circles, labeled by $4,2_{L},2_{R}$, correspond to the $U(4), Sp(2)_{L}, Sp(2)_{R}$ gauge groups, respectively. 
The $U(4)$ stack is identified with its mirror image through an $\Omega^{+}$-plane.  
$Sp(2)_{L,R}$ correspond to stacks of two D6-branes lying on the $\Omega^{+}$-plane. 
The triangles are $E2$-branes lying on the $\Omega^{+}$ plane, 
corresponding to $O(1)$ instantons. 
$E2', E2''$-instantons generate a quartic superpotential for $\Delta(10,1,1)$ and $\Delta^{c}(10,1,1)$, leading to an effective Majorana mass for the neutron. 
On the right, the effective unoriented quiver theory after Higgsing via $H, \bar{H}$
is shown. 
From the quiver on the left to the one on the right, extra undesired modulini appear, that are assumed to be
lifted by a combination of higgsing and fluxes. 
The $E2$-instanton generates a PMNS mass matrix for neutrinos.
The PS-like quiver generates the (MS)SM-like quiver on the right side after splitting the $Sp(2)_R$ D-branes from the $\Omega^{+}$-plane.
}
\label{plot}   
\end{figure}

The effective theory, in the low energy limit,
is described by a Pati-Salam gauge group $U(4){\times}  Sp(2)_{L}{\times}  Sp(2)_{R}$.
$U(4)$ is generated by a stacks of 4 D6-branes and their images
$U'(4)$ under $\Omega$ 
\footnote{Let us recall that $\Omega$-planes are introduced for 
quantum consistency and 
tadpole cancellations. See references \cite{Sagnotti0,Sagnotti1,Sagnotti2,Sagnotti3,Bianchi:1990yu, Bianchi:1990tb, Bianchi:1991eu, sessantatre, sessantaquattro, MBJFM,Bianchi:1990yu, Bianchi:1990tb, Bianchi:1991eu}
for a complete discussion of these aspects.}.
$Sp(2)_{L,R}$ are supported on two stacks of two D-branes each
lying on top of the $\Omega$-plane
\footnote{Let us note that, generically, in D-brane models,
one cannot construct directly $SU(N)$ gauge groups. 
For this reason we cannot obtain directly 
a PS model, but an extended one, 
with $U(4)$ rather than $SU(4)$ and 
$Sp(2)_{L,R}$ rather than $SU(2)_{L,R}$.
In fact, $N$ parallel branes stacked together 
(with open strings ending on them) 
will produce, at low energy limit, $U(N),SO(N),Sp(2N)$
gauge theories. In particular, 
 $U(N)$ is obtained if
 the D-brane stack
does not lie on the  
 $\Omega$-plane.
On the other hand, 
if the D-brane stack
lies on the $\Omega$-plane, 
one obtains $SO(N)$ or $Sp(2N)$
 (for $\Omega^{\mp}$ respectively).
$\Omega$-planes seem necessary in order to produce 
 realistic gauge groups, in which chiral matter 
can be embedded \cite{Angelantonj:1996uy, Angelantonj:1996mw}.}.
  We also consider three Euclidean $D2$-branes (or $E2$-branes)
  on top of the $\Omega$-plane, corresponding to 
three Exotic $O(1)$ Instantons. Let us call these $E2,E2',E2''$.   
Quarks and leptons in Left and Right  fundamental representations $F_{L,R}\equiv ({4,2_L}), ({4^*,2_R})$, are reproduced as open strings stretching from the $U(4)$-stack to the Left or Right $Sp(2)_{L,R}$-stacks (respectively). 
 Analogously, but at variant w.r.t. the original model \cite{Addazi:2015hka}, 
 Higgs $\bar{H} = ({4^*,2_R})$ and its conjugate $H = ({4,2_R})$ are introduced as 
 extra intersections of the $U(4)$-stack with $Sp(2)_{R}$.
 Extra color states $\Delta=(10,1,1)$, and their conjugates, are obtained as open strings stretching from the $U(4)$-stack to its $\Omega$ image $U(4)'$-stack. 
 $\phi_{LL}=(1,3,1)$ and $\phi_{RR}=(3,1,1)$ correspond to
strings with both end-points attached to the $Sp(2)_{L,R}$ (respectively).
 Higgs fields $h_{LR}=(2,2,1)$ are massless strings stretching from $Sp(2)_{L}$ to $Sp(2)_{R}$. 
 The quiver on the left of Fig.~1 automatically encodes the following super-potential terms
\cite{Addazi:2015hka}:
 \be{WYuk}
 \mathcal{W}_{Yuk}=Y^{(0)}h_{LR} F_{L}F_{R}+\frac{Y^{(1)}}{M_{F1}}F_{L}\phi_{LL}F_{L}\Delta+\frac{Y^{(2)}}{M_{F2}}F_{R}\phi_{RR}F_{R}\Delta^{c}
 \ee
 $$+\frac{Y^{(3)}}{M_{F3}}h_{LR}\phi_{RR}h_{RL}\phi_{LL}+\mu h_{LR}h_{RL}+Y^{(5)}h_{LR}F_{L}\bar{H}+\frac{Y^{(6)}}{M_{F6}}F_{R}\phi_{RR}\bar{H}\Delta^{c}$$
  $$+\frac{Y^{(7)}}{M_{F7}}F_{L}F_{L}F_{R}F_{R}+\frac{Y^{(8)}}{M_{F8}}F_{L}F_{L}\bar{H}\bar{H}+\frac{Y^{(9)}}{M_{F9}}F_{L}F_{L}F_{R}\bar{H}$$
  \be{WH}
 \mathcal{W}_{H}=m_{\Delta}\Delta\Delta^{c}+\frac{1}{4 M_{F4}}(\Delta\Delta^{c})^{2}+\frac{1}{2}m_{L}\phi_{LL}^{2}+\frac{1}{2}m_{R}\phi_{RR}^{2}+\frac{1}{3!}a_{L}\phi_{LL}^{3}
 \ee
 $$+\frac{1}{3!}a_{R}\phi_{RR}^{3}+\mu'H\bar{H}+\mu''F_{R}H+\frac{Y^{(10)}}{M_{F10}}\bar{H}\phi_{RR}\bar{H}\Delta^{c}$$
 \be{WE2}
\mathcal{W}_{E2',E2''} =\frac{Y^{'(1)}}{\mathcal{M}'_{0}}\epsilon^{ijkl}\epsilon^{i'j'k'l'}\Delta^{c}_{ii'}\Delta^{c}_{jj'}\Delta^{c}_{kk'}\Delta^{c}_{ll'}+\frac{Y^{''(1)}}{\mathcal{M}''_{0}}\epsilon^{ijkl}\epsilon^{i'j'k'l'}\Delta_{ii'}\Delta_{jj'}\Delta_{kk'}\Delta_{ll'}
\ee
$Y^{(...)}$ are $3{\times}  3$ Yukawa matrices;
the mass scales $M_{F...}$ are considered as free parameters:
 they depend on the particular completion of our model, {\it i.e.}
they could be near $M_{S}$,  the string scale, as well as at lower scales\footnote{The mass terms $m_{\Delta}$ and $m_{L,R}$ can be generated
by R-R or NS-NS 3-forms fluxes in the bulk,
in a T-dual Type IIB description, {\it i.e}  
$m_{\Delta}\sim\Gamma^{ijk}\langle\tau H_{ijk}+iF_{ijk}\rangle$,
$m_{L,R}\sim\Gamma^{ijk}\langle\tau H^{(L,R)}_{ijk} + iF^{(L,R)}_{ijk}~\rangle$,
with $H_3$ RR-RR and $F_3$ NS-NS 3-forms.
In general, $H_3,F_3$ are not flavour diagonal since fluxes through different cycles, wrapped by different D-branes, could be different. For recent discussions of 
mass deformed quivers and dimers see \cite{Bianchi:2014qma}.}.
 The super-potential terms (\ref{WE2})
can be generated by two $E2$-brane instantons
shown in Fig.1:  $O(1)', O(1)''$ intersect twice the $U(4)$ stack 
and $O(1)$ intersects twice the $U(4)$-stack and once the $Sp(2)_{R}$-stack ($2_{R}$ on the left side of Fig.1). 
In fact, fermionic modulini $\tau_{i},\tau'_{i},\omega'_{\alpha}$
appear as massless excitations 
of open strings ending on $U(4){-}O(1)$, $U(4){-}O(1)'$,
$Sp(2)_{R}{-}O(1)'$ respectively;
$i=1,4$ and $\alpha=1,2$ are indices of $U(4)$ and $Sp(2)_{R}$ respectively. 
Integrating over the fermionic modulini, we exactly 
recover the interactions (\ref{WE2p})
and (\ref{WE2}), as shown in \cite{Addazi:2015hka} or in  \cite{Blu1,Ibanez1,Ibanez2,Blu2} in different contexts\footnote{In \cite{Blu1,Ibanez1,Ibanez2,Blu2} Majorana masses for neutrinos are completely generated by exotic instantons.}. 
The dynamical scales generated in 
(\ref{WE2})
are 
 $\mathcal{M}_{0}'=Y^{'(1)}M_{S}e^{+S_{E2'}}$
 and  $\mathcal{M}_{0}''=Y^{''(1)}M_{S}e^{+S_{E2''}}$,
 where 
$S_{E2',E2''}$ depend on geometric moduli, associated to 3-cycles of the $CY_{3}$, around which $E2',E2''$ 
are wrapped.

The spontaneous breaking pattern down to the (MS)SM (minimal supersymmetric standard model) is
\be{pattern}
U(4){{\times} }Sp(2)_{L}{{\times} }Sp(2)_{R} \underset{\langle {\rm Stu}  \rangle}  {\longrightarrow} SU(4){{\times} }Sp(2)_{L}{{\times} }Sp(2)_{R}
\ee
$$ \underset{\langle \bar{H},H,h \rangle}   {\longrightarrow} SU(3) {\times}  Sp(2)_{L}{\times}  U(1)_{Y}$$
(Stu stands for St\"uckelberg, see below) and $h_{LR}$ contain the standard Higgses for the final electroweak symmetry breaking.
Decuplets decompose as
$\Delta^{c}=\Delta_{6}^{c}+T^{c}+S^{c}$, 
 with $\Delta_{6}={6}_{Y=+2/3}$,
 $T={3}_{Y=-2/3}$, $S={1}_{Y=-2}$,
 and the singlet $S$ takes a VEV.

Let us note that the extra $U(1)_{4}\subset U(4)_{c}$ 
is anomalous in gauge theory. In string theory  
a generalization of the Green-Schwarz mechanism can cure 
these anomalies. Generalized Chern-Simons 
(GCS) terms are generally required in this mechanism. 
The new vector boson $Z'$ associated to $U(1)_{4}$
 gets a mass via a St\"uckelberg mechanism\footnote{ See  \cite{Stuck1,Stuck2,Stuck3,Stuck4,Stuck5,Stuck6,Stuck7,Stuck8,Stuck9,Stuck10,Stuck11,Stuck12} for discussions about these aspects in different contexts, and  \cite{Addazi:2015hka} for comments on implications in PS models, like $Z_{R}-Z'$ mixings or GCS interactions $Z_{R}-Z'-Z$ or $Z_{R}-Z'-\gamma$ etc (where $Z_{R}$ is the $SU(2)_R$ Z-boson). 
Another implementation of
 the St\"uckelberg mechanism is in the realization of Lorentz Violating
 Massive gravity \cite{LIV1,LIV2,LIV3}. Recently, geodetic instabilities of St\"uckelberg
 Lorentz Violating
 Massive gravity were discussed in \cite{Addazi:2014mga} (and also connected to solutions of naked singularities discussed in \cite{Addazi:2015gna}). We would like to stress that GCS terms 
generate UV divergent triangles that are cured by considering UV completions with KK states or string excitations.
For issues in scattering amplitudes and collider physics see \cite{Santini}.
See also 
\cite{Addazi:2015dxa,Addazi:2015ppa}
for a string-inspired non-local field model of string theory.}.

The final effective (MS)SM embedding quiver  that we will consider is 
obtained from the previous SUSY PS-like quiver through a splitting of nodes
$4\rightarrow 3+1$ and $2_{R}\rightarrow 1+1'$.  
 In this new quiver, $E2$ intersects $U(1)$ and $\hat{U}(1)'$ as shown on the right of  Fig.~1,
where $\hat{U}'(1)$ is the $\Omega$-image of $U'(1)$. 
In the Higgsing from SUSY PS-like quiver to SUSY SM-like, 
extra undesired modulini are obtained. In particular, colored modulini at $E2{-}U(3)$ intersections. 
We assume that these modulini are lifted out by Higgsings and fluxes. 
This technical aspect deserves future investigation beyond the purposes of this paper. 
As a consequence, an extra mass matrix term is non-perturbatively generated 
\be{WE2p}
\mathcal{W}_{E2}= {1\over 2} \mathcal{M}_{ab}'N^{a}_{R}N^{b}_{R}
\ee
where $N_{R}^{a}$ are RH neutrinos  ($a=1,2,3$ label neutrino species), contained, as singlet, inside $F_{R}$. 
The generated mass matrix is
$\mathcal{M}_{ab}=Y_{ab}^{(0)'}M_{S}e^{-S_{E2}}$, where
$Y_{ab}^{(0)'}$ is the Yukawa matrix 
parameterizing masses and mixings among RH neutrinos,
depending of course on the particular $E2$ intersections 
with ordinary D6-branes stacks.
Let us note that the superpotential (\ref{WE2p}) can be generated only after 
spontaneous symmetry breaking of $U(4)_{c}$ down to 
$U(3)_{c}$, and $Sp(2)_{R}$ down to $U'(1)$.
This will impose bounds on the parameters
that we will discuss in Section 2. 

Now, let us discuss electroweak symmetry breaking in our present model:
as mentioned  before, this is due to the VEVs $\langle h_{LR} \rangle$ of the complex Higgs bi-doublets 
$h_{LR}$
yielding the tree-level mass relations for leptons and quarks 
\be{treelevel}
m_{d}=m_{e}\,\,\,\,\,{\rm and}\,\,\,\,\,\,m_{u}=m_{D}
\ee
 where $m_{D}$ are Dirac masses of neutrinos.
From (\ref{treelevel}), tight hierarchy constraints on RH neutrino  masses
are predicted: as a result the neutrino's hierarchy is related to the
up-quarks. It is interesting to observe that the hierarchy obtained at the perturbative level
(with closed-string fluxes generating the $M_{2}$ scale)
is corrected by exotic instantons, 
parametrized by $\mathcal{M}_{ab}$.  
Left-Right symmetry breaking pattern implies 
\be{wh}
m_{D}=m_{u}\,\,\,\,{\rm and}\,\,\,\,\,\,V_{L}=V_{CKM}
\ee
with $V_{CKM}$ the Cabibbo-Kobayashi-Maskawa matrix.
We obtain the mass matrix
\be{MassMatrix}
M= \left( \begin{array}{cc} 0 & m_{D}
\ \\ m_{D} & M_{R} \ \\
\end{array} \right) 
\ee
In our case, RH neutrino masses are 
$$M_{R}=M_{R}^{P}+M_{R}^{E2'}$$
where 
$$M_{R}^{P} =\langle\phi_{RR}\rangle\langle S^{c} \rangle/M_{2}$$
and 
$$M_{R}^{E2}=\mathcal{M}_{ab}'$$ 
as shown in \cite{Addazi:2015hka}.

From the usual see-saw formula one obtains the light neutrino mass matrix $m_{\nu}$
\be{ssf}
m_{\nu}\simeq -m_{D}\left(M_{R}^{P}+M_{R}^{E2}\right)^{-1}m_{D}
\ee
A natural situation for our quiver is that 
$E2'$ induce non-perturbative mass terms for RH 
neutrinos  of the same order, {\it i.e.} $M^{E2}_{R,1}\simeq M^{E2}_{R,2} \simeq M^{E2}_{R,3}$ where 1,2,3 are generation indices. As a consequence, 
$M^{E2}_{R,1,2,3}\simeq 10^{9\div 13}\ \rm GeV$ and
we obtain a highly  degenerate RH mass spectrum
in a good range for leptogenesis, non-perturbative mass corrections are higher than or at least of the same order as
the perturbative ones. 
Naturally, such a situation does not imply a highly degenerate LH mass spectrum, since a large quark-lepton hierarchy 
remains encoded in $m_{D}$. 
The see-saw formula can be inverted as
\be{issf}
M_{R}=M_{R}^{p}+M_{R}^{E2}\simeq -m_{D}m_{\nu}^{-1}m_{D}
\ee
since in our model $m_{D}=m_{D}^{T}$.
From (\ref{issf}) one can get information 
on the RH neutrino mass matrix $M_{R}$
by using data on LH neutrino mass matrix
$m_{\nu}$, and assuming a quark-lepton symmetry.
In general, a quark-lepton symmetry 
complicates BAU mechanisms 
because it imposes a strong hierarchy in the neutrino sector:
under the assumption that $v_{1}v_{2}/M_{2}\simeq 10^{11\div 13}\,\rm GeV$
with $v_{1}=\langle \phi_{RR} \rangle$ and $v_{2}=\langle \Delta^{c} \rangle$, the lightest RH eigenstate $N_{1}$ takes a mass 
much smaller than the Davidson-Ibarra bound \cite{DI}, $M_{N_{1}} \ll 10^{9}\, \rm GeV$, {\it i.e} $N_{1}$ decays cannot guarantee a sufficient production 
of lepton asymmetry.
Fortunately, non perturbative $E2$ contributions 
can generate a
 compact RH neutrino spectrum above the DI bound,
 {\it i.e.} the mass eigenvalues of RH neutrino mass 
 matrix are highly degenerate and higher than $10^{9}\, \rm GeV$.
 We would like to stress that, unlike 
 $SO(10)$ GUTs,  our model provides a 
natural mechanism to  obtain a compact RH neutrino hierarchy.
 Let us also observe that, after the splitting in Fig.~1, we  obtain an effective cubic interaction term 
$(\langle S^{c} \rangle/\mathcal{M}_{0})\epsilon^{SU(3)}_{ijk}\epsilon^{SU(3)}_{i'j'k'}\Delta_{6}^{c^{ii'}}\Delta_{6}^{c^{jj'}}\Delta_{6}^{c^{kk'}}$
which violates Baryon number as $\Delta B=2$ and
generates a Majorana mass for neutrons \cite{Addazi:2015hka}, as mentioned in the introduction. On the other hand, exotic instantons 
can preserve discrete sub-symmetries $Z_{2}^{(\Delta B,\Delta L=1)}$,
avoiding proton destabilization, but allowing $\Delta L,\Delta B=\pm 2$ processes. 
However, $\Delta B=2$ violating operators can also destabilize the proton if one consider 
all $\Delta L=1$  mixing terms among $F_{L,R}$ and $H,\bar{H}$ in (\ref{WYuk}). 
Higher order operators of this kind are 
generated by fluxes, so that one can naturally 
assume that they are suppressed by a mass scale larger than  
$M_{S}$. 

 So, potentially dangerous 
terms are $$
{\cal W}_{Y^{(5)},\Delta L=1}=Y^{(5)}h_{LR}F_{L}\langle \bar{H} \rangle \quad 
{\rm and} \quad {\cal W}_{\mu,\Delta L=1} = \mu''F_{R}H $$
These terms are easily understood: 
$\bar{H}$ is like a fourth generation of $F_{R}$.
So that, calling $\mathcal{F}_{R}^{f=1,4}=(F_{R}^{f=1,3},\bar{H})$, 
they generically mix through 
$$\mathcal{W}_{\mu_{f}}=\mu_{f}H\mathcal{F}_{R}^{f}=\mu' H\bar{H}+\mu''HF_{R}$$
Such mass terms can be diagonalized so that the mixing term $HF_{R}$
can be rotated away in the mass eigenstate basis. 
Similarly, $\mathcal{W}_{Y^{(5)}}$ can be incorporated in the standard Yukawa term 
as
$$\mathcal{W}_{\mathcal{Y}^{f}}=\mathcal{Y}_{f'=1,3;f=1,4}h_{LR}F_{L}^{f'}\mathcal{F}_{R}^{f}=h_{LR}\left[Y^{(0)}_{f'=1,3,f=1,3}F_{L}^{f'}F_{R}^{f}+Y^{(5)}_{f'=1,3}F_{L}^{f'}\bar{H}\right]$$
In order to avoid proton destabilization, we can impose the following condition on matrices $\mu_{f}$ and $\mathcal{Y}^{f}$
\be{muY}
\mu_{f}\mathcal{Y}^{f}=0
\ee
Relation (\ref{muY}) automatically guarantees matrices of the form
$$\mu_{f}=(\mu',0,0,0)^{T}$$
$$\mathcal{Y}_{f}=(0,Y^{(0)}_{f=1},Y^{(0)}_{f=2},Y^{(0)}_{f=3})$$
in the basis $\mathcal{F}^{f}_{R}=(F_{R}^{f=1,2,3},\bar{H})$. 

A natural geometric explanation of Eq.(\ref{muY}) could come from global intersecting D-brane models,
consistently completing our local one in the Calabi-Yau singularity. 
The quiver in Fig.~1 apparently seems to {\it democratically} consider different flavors,  
like $\mathcal{F}^{f}_{R}$. However,
the presence of internal bulk R-R  or NS-NS fluxes 
can discriminate different intersections of two stacks of D6-branes
{\it i.e.} different flavors from one another. 
Alternatively, one can consider that the local quiver theory in Fig.~1 
could come from a (or more) Gepner model(s). 
In Gepner models, the Calabi-Yau space has a more complicated geometry than for example a toroidal orbifold, inducing 
accidental discrete symmetries in the low energy limits. 
For example, the intersections of two stacks on a torus are geometrically equivalent, while in a complicated topological deformation of a torus
``flavor democracy'' is broken.  This affects the vertex operators of 
an open string massless fermion $\mathcal{V}_F=\mathcal{V}_{S}\Sigma^{int}_{f}$, 
where $\mathcal{V}_{S}=u^{\alpha}(k)S_{\alpha}e^{-\phi/2}e^{ikX}$ accounts for the space-time part, while $\Sigma^{int}_{f}$ is an internal spin field depending on the flavor. Similarly for massless scalars $\mathcal{V}_{B} = \Psi^{int}_{f} e^{-\phi} e^{ikX}$, with $\Psi^{int}_{f}$ being a chiral primary operator. A Yukawa coupling, like $h_{LR} F_{L}\mathcal{F}_{R}$, will give rise to a flavor matrix $Y_{f_{1}f_{2}f_{3}}$
proportional to  $\langle \Psi^{int}_{f_1} \Sigma^{int}_{f_2}\Sigma^{int}_{f_3} \rangle$. As a consequence, the suppression 
of $\mathcal{W}_{Y^{(5)},\Delta L=1}$ can be geometrically understood as
emerging from different inequivalent intersections among the same stacks of branes \footnote{
For recent literature on emergent discrete symmetries in (MS)SM-like and PS-like models,
see \cite{Discrete1,Discrete26} and references therein.}.

\subsection{Free parameters}
\label{freepar}
In this section we will comment on the relevant parameters in our model and clarify our 
assumptions. 

\subsubsection{Supersymmetry and string scale}
First, let us  clarify the role of supersymmetry in our considerations. 
Clearly, if the SUSY breaking scale is assumed to be $M_{SUSY}\simeq 1\, \rm TeV$, 
this will introduce several extra parameters relevant for leptogenesis.
A TeV-scale SUSY will complicate one-loop (n-loops) contributions,
introducing extra CP-violating phases in RH-neutrino  decays.
Here, we will assume that supersymmetry has nothing to 
do with the hierarchy problem of the Higgs mass, 
{\it i.e.} SUSY has the role to stabilize instanton calculations 
and to eliminate tachyonic states from the present string model. 
While the second aspect is crucial for the consistency 
of our model, saving us from ``fighting" with instabilities, 
and imposing a bound on the SUSY-scale as $M_{SUSY}\simeq M_{S}$, 
the first aspect is ``less fundamental", 
since it only has the role of simplifying istanton calculations. 
This requires $M_{SUSY}\simeq M_{S}e^{-S_{E2}}\gtrsim 10^{9}\, \rm GeV$.
As a result, supersymmetric particles do not give any relevant contributions to 
RH neutrino  decays\footnote{One could speculate that dark matter is a hidden parallel system of intersecting D-branes.
Implications in direct detection of such a scenario was studied in \cite{Addazi:2015cua}. }.

\subsubsection{Relevant effective Lagrangian and free parameters}
\label{freepar1}

After the spontaneous breaking of 
SUSY, $U(4)$ symmetry and Left-Right symmetry,
the effective Lagrangian in the neutrino  sector reads
\be{relevant}
\mathcal{L}_{eff}^{\nu}=Y^{(0)}\langle h_{u}\rangle l \nu_{R}+
\frac{Y^{(2)}}{M_{2}}\nu_{R}\langle \varphi_{RR} \rangle \nu_{R}\langle \delta^{c}\rangle+
Y^{(0)'}M_{S}e^{-S_{E2}}\nu_{R}\nu_{R}
\ee
where $h_{u}$ is the scalar component 
of the superfield $H_{u}$ contained in the bi-doublet superfield $h_{LR}$, 
$\nu_{R}$ are the RH neutrinos,
the fermionic component of the  
the RH neutrino  supermultiplets,
$\varphi_{RR},\delta^{c}$ are the scalar components 
of the supermultiplets $\phi_{RR},\Delta^{c}$.

Therefore,  the number of relevant free parameters in the neutrino sector
is  
\be{number}
N_{f.p.}=n_{Y0}+n_{Y2}+n_{Y0'}+n_{VEV1}+n_{VEV2}+n_{Flux}+n_{E2}=22
\ee
($f.p.$ stands for free-parameters)
where $n_{Y0,2}=6$ are the number of free parameters 
in the Yukawa matrices $Y^{(0)},Y^{(2)},Y^{(0)'}$ respectively;
$n_{VEV1,VEV2}$ account for the number of ratios between extra VEVs
$v_{1,2}$ with respect to $v_{EW}$, {\it i.e.} $z_{1}=v_{1}/v_{EW}$
and $z_{2}=v_{2}/v_{EW}$; $n_{Flux}=1$ is the number of non-perturbative scales generated by fluxes entering in the neutrino sector, 
{\it i.e} $M_{F2}$ (or $z_{3}=M_{F2}/v_{EW}$); $n_{E2}$ parameterizes the size of the 3-cycle wrapped by 
$E2$-brane.

Under reasonable assumptions, the 
number of free parameters can be significantly reduced. 
In the following analysis, we will 
suppose a dominance of non-perturbative effects: 
$M_{R}^{E2} \gg M_{R}^{P}$ (all matrix parameters). 
In this case, $n_{VEV1,VEV2,Flux,Y2}$ are irrelevant, 
as they are related 
to  tiny extra corrections. 
In this case, the mass matrix of RH neutrinos  is 
practically completely generated by the $E2$-instanton!
 AB: The hierarchy $M_{R}^{E2} \gg M_{R}^{P}$ can be understood as follows.
The $E2$-instanton generates a mass matrix
for neutrinos with an absolute value $M_{S}e^{-\Pi_{3}/g_{s}}$, 
where $\Pi_{3}$ is the volume of 3-cycles wrapped by the $E2$-instanton
on $CY_{3}$. Volumes of 3-cycles (in string units) can be as small as $\Pi_{3}\simeq 1$, 
or as large as $\Pi_{3}>>1$. In other words, the
hierarchy among RH neutrino masses and the string scale can be considered as a free parameter.
On the other hand, the $Y^{(2)}$-term is  suppressed by the scale of the non-perturbative flux, that can easily be near the string-scale so as to justify the assumed hierarchy $M_{R}^{E2} \gg M_{R}^{P}$. 

As a consequence, the number of relevant parameters will simply be 
\be{number2}
N_{f.p.}\simeq n_{Y0}+n_{Y0'}+n_{E2}=6+6+1=13
\ee
Let us note that such a situation requires
$v_{1}v_{2}/M_{F2} \ll 10^{9}\, \rm GeV$.
But $v_{1,2}<v_{R}$ with $v_{R}\gtrsim 10^{9}\, \rm GeV$:
exotic instanton effects are related to a St\"uckelberg mechanism 
for $U(1)_{B-L}$, otherwise they will violate the B-L gauge symmetry.
On the other hand, $v_{R}\gtrsim 10^{9}\, \rm GeV$
since exotic instantons have to distinguish RH neutrinos  
from $E^{c}$ at this very scale!
As a consequence, $M_{F2} \gg 10^{9}\, \rm GeV$  satisfies
these bounds. This situation seems natural: $M_{F2}$ are related to closed-string fluxes,  
{\it i.e.} another kind of quantum gravity effects.

\section{ Phenomenology in neutrino physics}
In this section we derive our predictions for yet-unknown low energy neutrino parameters, the mass of the lowest neutrino state and the phases of the   PMNS (Pontecorvo-Maki-Nakagawa-Sakata) matrix.

\subsection{Conditions for a compact RH neutrino spectrum}
\label{CompactRH}

As mentioned in Sect. \ref{PSlikeABDbrane}, the  Dirac neutrino mass matrix $m_{D}$ is symmetric, thus it can be diagonalized
 by a single unitary matrix 
$V_{L}$ \cite{Takagi, Pontecorvo:1957cp, Pontecorvo:1967fh}
\be{p1}
m_{D}=V_{L}^{\dagger}m_{D}^{diag}V_{L}^\ast
\ee
where  $m_{D}^{diag} \equiv diag(m_{D1},m_{D2},m_{D3})$
with  real and non-negative eigenvalues  $m_{(D1,D2, D3)}$.
The seesaw condition expressed in  Eq.  (\ref{issf}) yields
\be{MR2}
M_{R}=-V_{L}^{\dagger}m_{D}^{diag}Am_{D}^{diag}V_{L}^{*}
\ee
where we have defined a matrix $A$, symmetric by construction, as
\be{A}
A=V_{L}^{*}m_{\nu}^{-1}V_{L}^{\dagger}
\ee
In terms of the matrix elements of $A$ and $V_L$, the RH mass matrix elements  become
\begin{eqnarray}
\label{RHmm1}
M_{R11}&=& 
-A_{11}V_{L11}^{\ast2} m_{D1}^{2}-A_{22}V_{L21}^{\ast2} m_{D2}^{2}-A_{33}V_{L31}^{\ast2} m_{D3}^{2}+
 \nonumber \\
&-& 2 A_{12} V_{L11}^\ast  V_{L21}^\ast m_{D1} m_{D2}-2 A_{13}  V_{L11}^\ast   V_{L31}^\ast m_{D1}m_{D3} -2 A_{23}  V_{L11}^\ast   V_{L21}^\ast m_{D2}m_{D3} \nonumber 
\end{eqnarray}
\begin{eqnarray}
M_{R12}&=& 
-A_{11}V_{L11}^{\ast} V_{L12}^{\ast}m_{D1}^{2}-A_{22}V_{L21}^{\ast} V_{L22}^{\ast} m_{D2}^{2}-A_{33}V_{L31}^{\ast}V_{L32}^{\ast} m_{D3}^{2}+
 \nonumber \\
&-&  A_{12} (V_{L12}^\ast  V_{L21}^\ast+ V_{L11}^\ast  V_{L22}^\ast) m_{D1} m_{D2}- A_{13}  (V_{L12}^\ast   V_{L31}^\ast+V_{L11}^\ast   V_{L32}^\ast) m_{D1}m_{D3} +
 \nonumber \\
&-&   A_{23}  (V_{L22}^\ast   V_{L31}^\ast+V_{L21}^\ast   V_{L32}^\ast) m_{D2}m_{D3} \nonumber
\end{eqnarray}
\begin{eqnarray}
M_{R13} &=& 
-A_{11}V_{L11}^{\ast} V_{L13}^{\ast}m_{D1}^{2}-A_{22}V_{L21}^{\ast} V_{L23}^{\ast} m_{D2}^{2}-A_{33}V_{L31}^{\ast}V_{L33}^{\ast} m_{D3}^{2}+
 \nonumber \\
&-&  A_{12} (V_{L13}^\ast  V_{L21}^\ast+ V_{L11}^\ast  V_{L23}^\ast) m_{D1} m_{D2}- A_{13}  (V_{L13}^\ast   V_{L31}^\ast+V_{L11}^\ast   V_{L33}^\ast) m_{D1}m_{D3} +
 \nonumber \\
&-&   A_{23}  (V_{L23}^\ast   V_{L31}^\ast+V_{L21}^\ast   V_{L33}^\ast) m_{D2}m_{D3} \nonumber 
\end{eqnarray}
\begin{eqnarray}
M_{R22}&=& 
-A_{11}V_{L12}^{\ast2} m_{D1}^{2}-A_{22}V_{L22}^{\ast2} m_{D2}^{2}-A_{33}V_{L32}^{\ast2} m_{D3}^{2}+
 \nonumber \\
&-& 2 A_{12} V_{L12}^\ast  V_{L22}^\ast m_{D1} m_{D2}-2 A_{13}  V_{L12}^\ast   V_{L32}^\ast m_{D1}m_{D3} -2 A_{23}  V_{L22}^\ast   V_{L32}^\ast m_{D2}m_{D3} \nonumber 
\end{eqnarray}
\begin{eqnarray}
M_{R23}&=& 
-A_{11}V_{L12}^{\ast} V_{L13}^{\ast}m_{D1}^{2}-A_{22}V_{L22}^{\ast} V_{L23}^{\ast} m_{D2}^{2}-A_{33}V_{L32}^{\ast}V_{L33}^{\ast} m_{D3}^{2}+
 \nonumber \\
&-&  A_{12} (V_{L13}^\ast  V_{L22}^\ast+ V_{L12}^\ast  V_{L23}^\ast) m_{D1} m_{D2}- A_{13}  (V_{L13}^\ast   V_{L32}^\ast+V_{L12}^\ast   V_{L33}^\ast) m_{D1}m_{D3} + \nonumber \\
&-&   A_{23}  (V_{L23}^\ast   V_{L32}^\ast+V_{L22}^\ast   V_{L33}^\ast) m_{D2}m_{D3} \nonumber 
\end{eqnarray}
\begin{eqnarray}
M_{R33}&=& - A_{11}V_{L13}^{\ast} m_{D1}^{2}-A_{22}V_{L23}^{\ast2} m_{D2}^{2}-A_{33}V_{L33}^{\ast2} m_{D3}^{2}- 2 A_{12} V_{L13}^\ast  V_{L23}^\ast m_{D1} m_{D2} + \nonumber \\
&-& 2 A_{13}  V_{L13}^\ast   V_{L33}^\ast m_{D1}m_{D3}-2 A_{23}  V_{L23}^\ast   V_{L33}^\ast m_{D2}m_{D3}
\end{eqnarray}
Since the matrix $M_R$ is also symmetric by construction, one has $M_{Rij}=M_{Rji}$ for any $i,j ={1,2,3}$.
 Motivated by  quark-lepton symmetry, we assume, as for quarks,  a large hierarchy in
the eigenvalues of the Dirac mass matrix for leptons, that is
\be{hm}
m_{D1} \ll m_{D2} \ll m_{D3}
\ee
The hierarchy assumption in \eqref{hm} implies that the elements of $A$ are at most mildly hierarchical, and the same holds for the RH neutrino spectrum.
Therefore only specific constraints on  the $A$ matrix can enforce the conditions
that ensure that the RH neutrino spectrum is compact. We can immediately see that a generically compact RH spectrum would result by suppressing the entries proportional to $A_{23}$ and  $A_{33}$. In that case, all  matrix elements become of the same order of magnitude, that is $m_{D1} m_{D3}
\sim m_{D2}^2$.
In first approximation, we can set
\begin{equation}
  \label{eq:conditions}
 A_{23} =A_{33}=0\,.
\end{equation}
Let us stress that while the approximation \eqref{eq:conditions} has the
virtue of simplifying the analysis, a generic compact RH neutrino spectrum can be obtained by fixing the $A_{23}$ and $A_{33}$
values to any sufficiently small number.

The precise form of the $V_L$ matrix is not crucial to ensure the compactness of the RH spectrum, provided it does not have  unnaturally large matrix elements.  
Guided by the  symmetries of the model,  discussed in  Sect. \ref{PSlikeABDbrane}, we assume 
that in  the diagonal basis for the down-quarks and charged leptons mass matrices,
the unitary rotation $V_L$ that diagonalizes the symmetric matrix $m_D$ coincides with the Cabibbo-Kobayashi-Maskawa
(CKM) matrix that diagonalizes $m_u$. In other terms, we set, according to Eq. \eqref{wh}
\be{hm1}
V_L = V_{CKM}
\ee
where $ V_{CKM}$ is the CKM matrix encoding quark mixing.


\subsection{Low Energy Observables}
\label{LEO}

The   PMNS matrix is 
 the lepton conterpart of the CKM mixing matrix in the quark sector. If neutrinos  are Majorana particles,   there are two more physical phases with respect to the CKM matrix.
By adopting  the standard parametrization  in terms of  three Euler mixing angles $\theta_{12}$, $\theta_{23}$ and $\theta_{13}$,
 a Dirac phase $\delta$, and two Majorana phases $\alpha$ and $\beta$, the  PMNS  mixing matrix can be written as:
\begin{eqnarray}
\label{eq:UPMNS}
U_{PMNS} & = & U_{PMNS}^\prime(\theta_{12}, \theta_{23}, \theta_{13}, \delta)
{\times} {\rm diag}\left(1, e^{i\alpha}, e^{i\beta} \right)\,.
\end{eqnarray}
where 
\begin{eqnarray}
\label{eq:UPMNS1}
U_{PMNS}^\prime & = & \left(\begin{array}{ccc}
c_{12}c_{13} & s_{12}c_{13} & s_{13}e^{-i\delta}\\
-s_{12}c_{23}-c_{12}s_{23}s_{13}e^{i\delta} & c_{12}c_{23}-s_{12}s_{23}s_{13}e^{i\delta} & s_{23}c_{13}\\
s_{12}s_{23}-c_{12}c_{23}s_{13}e^{i\delta} & -c_{12}s_{23}-s_{12}c_{23}s_{13}e^{i\delta} & c_{23}c_{13}\end{array}\right)
\end{eqnarray}
Here $c_{ij}=\cos \theta_{ij}$ and $s_{ij}=\sin \theta_{ij}$, with $i$
and $j$ labeling families that are coupled through that angle ($i, j =
1, 2, 3$).  
 In the basis in which the charged lepton mass matrix is diagonal, $U_{PMNS}$
diagonalizes the effective neutrino mass matrix
\be{MLL}
m_\nu = U_{PMNS}^* m_\nu^{diag} U_{PMNS}^\dagger
\ee
where
\be{mdiag1}
m_\nu^{diag} = {\rm diag}(m_1,m_2,m_3)
\ee
Since the matrix $V_L$ is also unitary, we  choose  the same parameterization   as for the PMNS matrix, Eq. \eqref{eq:UPMNS1},   distinguishing the $V_L$ parameters 
 with a  prime superscript:
$s_{12}',\,s_{23}',\,s_{13}',\,\delta'$. Their values are the same as the ones in the CKM matrix because of the assumption $V_L=V_{CKM}$,
  discussed in Sect. \ref{CompactRH}. 

In Sect. \ref{freepar1} we have operated 
a  counting of the fundamental free parameters of the model, and found  13 real parameters
in the case of dominance of non-perturbative effects.
Under the assumption of symmetry expressed by Eq. \eqref{wh}, 
the values of
these 13 real parameters are constrained by observables in the up-type quark and neutrino sectors.
They are: the three quark masses
$m_u,\,m_c,\,m_t$, the two neutrino mass-squared differences $\Delta
m^2_{21},\, \Delta m^2_{32}$, the three CKM mixing angles
$\theta_{12}',\,\theta_{23}',\,\theta_{13}'$ and the three PMNS mixing
angles $\theta_{12},\,\theta_{23},\,\theta_{13}$, amounting to 
 $11$ real observables. Imposing on the complex elements of
the matrix $A$ the  two additional conditions in \eqref{eq:conditions}, $A_{23}=A_{33}=0$,
implies that the set of  real
fundamental parameters must satisfy two additional requirements, that
is  ${\rm Re}(A_{23})={\rm Re}(A_{33})=0$.
%
Thus the parameter space of the model remains completely determined,
allowing to obtain a quantitative prediction for the absolute neutrino
mass scale $m_{1}$.


The matrix $A$  can  be expressed in terms
of the observables  $V_L$,  $U_{PMNS}$ and  $ m_\nu^{diag}$ as
\begin{equation}
  \label{eq:AVUm}
  A= \left(V_L U_{PMNS}^\ast \right)^\ast \, \frac{1}{m_\nu^{diag}}\,
\left(V_L  U_{PMNS}^\ast \right)^\dagger\,.
\end{equation}
This equality connects $A$ to the observables listed before, and  the conditions
$
A_{23}= A_{33}=0 
$
determine two relations among them, that we generically indicate with
\begin{eqnarray}
  \label{eq:delta1}
f([\theta^\prime_{ij}, \delta^\prime, \theta_{12},\theta_{23},\theta_{13}, \Delta m^2_{21}];\delta, m_1, \alpha, \beta)&=&0\\
  \label{eq:delta2}
g([
\theta^\prime_{ij}, \delta^\prime, \theta_{12},\theta_{23},\theta_{13}, \Delta m^2_{31}];\delta, m_1, \alpha, \beta) &=&0
\end{eqnarray}
where  $f$ and $g$ are  known  functions.
We  have eliminated
$m_2$ and $m_3$ by using their relations with 
their mass-squared differences,  $m^2_2=m^2_1+\Delta
m^2_{21}$ and $m^2_3=m^2_1+\Delta m^2_{31}$.
By projecting $f$ and $g$ onto their absolute values, we obtain two relations between real quantities connecting the mass $m_1$ and the PMNS phase $\delta$.
Extracting imaginary parts from equations \eqref{eq:delta1} and \eqref{eq:delta2} gives nontrivial relations between the observable $\delta'$ and the  PMNS phases, and allows to determine $\alpha$ and $\beta$ in terms of $m_1$, $\delta$, and the known mixing angles and mass squared differences.  

In Eqs. \eqref{eq:delta1} and \eqref{eq:delta2}  the input  parameters are listed  in square brackets. Their approximate averages, which for our purpose  represent an adequate level of approximation, are reported in
Table~\ref{tab:1}. 
%
\begin{table}[t!]
\begin{center}
\begin{tabular}{|c|c||c|c|}
  \hline
  \multicolumn{2}{|c||}{\rm Quark\ sector } &
  \multicolumn{2}{|c|}{\rm Neutrino\ sector }  \\ \hline
  &&& \\ [-8pt]
  $m_u(\Lambda)$ &$ 0.00067\> {\rm GeV}$&$ \Delta m^2_{21}(\Lambda)$&$ 11.71 {\times}  10^{-5}\> {\rm eV}^2$ \\  [2pt]
  $m_c(\Lambda)$ &$ 0.327\ \  {\rm GeV}$&$\Delta m^2_{31}(\Lambda)$&$ 3.84 {\times}  10^{-3}\> {\rm eV}^2$ \\  [2pt]
  $m_t(\Lambda)$ &$ 99.1\> \ \ \ \; {\rm GeV} $&   & \\  [4pt]
  \hline &&& \\ [-8pt]
  $\theta'_{12}$ &$ 13.03^\circ$ &$ \theta_{12}  $ &$ 33.5^\circ $\\  [2pt]
  $\theta'_{23}$ &$ \ \, 2.37^\circ $&$ \theta_{23}  $ &$ 42.3^\circ $\\  [2pt]
  $\theta'_{13}$ &$ \ \, 0.24^\circ$ &$ \theta_{13}  $ &$ \ 8.5^\circ$ \\  [2pt]
  $\delta'$     &$\quad\ \;  1.19\,$rad&    &  \\ [4pt]
  \hline
\end{tabular}
\caption{Input parameters. We use the up-quark masses renormalized to the
  scale $\Lambda = 10^{9}$ GeV given in Table IV in
  Ref.~\cite{Xing:2007fb}. neutrino's mass squared differences are
  taken from the global fit in Ref.~\cite{Gonzalez-Garcia:2014bfa} and
  renormalized to the scale $\Lambda$ with a multiplicative factor
  $r^2$ with $r=1.25$ according to the prescription in
  Ref.~\cite{Giudice:2003jh}.  The CKM mixing angles $\theta'_{ij}$ and CKM
  phase $\delta'$ are derived from the values of the Wolfenstein
  parameters given by the PDG~\cite{Agashe:2014kda}.  The PMNS mixing angles are
  taken from the global fit in
  Ref.~\cite{Gonzalez-Garcia:2014bfa}.
 Renormalization effects for the
  CKM and PMNS parameters have been neglected.}
\label{tab:1}
\end{center}
\vspace{-0.9cm}
\end{table}
 Neutrinos mass squared differences are
  taken from the global fit in Ref.~\cite{Gonzalez-Garcia:2014bfa} and
  renormalized to the scale $\Lambda = 10^{9}$  GeV ($\sim M_R$), with a multiplicative factor
  $r^2$ ($r=1.25$,  according to the prescription in
  Ref.~\cite{Giudice:2003jh}).
 The up-quark masses,   renormalized to the
  scale $\Lambda$, are  taken from Table IV in
  Ref.~\cite{Xing:2007fb}.
  The CKM mixing angles $\theta'_{ij}$ and CKM
  phase $\delta'$ are derived from the values of the Wolfenstein
  parameters given by the PDG~\cite{Agashe:2014kda}.  The PMNS mixing angles are
  taken from the global fit in Table 1 of
  Ref.~\cite{Gonzalez-Garcia:2014bfa}, under the assumption of normal hierarchy of the neutrino masses. 
 Renormalization effects for the
  CKM and PMNS parameters have been neglected.
It is worth noting
that the $|V_{ub}|$ puzzle keeps affecting the uncertainty of  the small  $\theta_{13}^\prime$ value\footnote{For reviews on  the $V_{ub}$ uncertainties  see {\it e.~g.} \cite{Ricciardi:2015iwa, Ricciardi:2014aya, Ricciardi:2014iga, Ricciardi:2013cda, Ricciardi:2013xaa, Ricciardi:2012pf}.}.

Given that the signs of $\theta_{12},\theta_{23}$
and $\theta_{13}$ are not determined in oscillation experiments,
depending on the possible choices $\pm\theta_{ij}$ the two
\refs{eq:delta1}{eq:delta2} represent in principle $2^3=8$ conditions.
We focus on the case $(\theta_{12}, \theta_{23})=(-|\theta_{12}|, -|\theta_{23}|)$, which, according to Ref. \cite{Buccella:2012kc}, where an analogous procedure is used  in the contest of non-SUSY $SO(10)$ GUT, is a phenomenologically acceptable case.

 The plots of $m_1$ as a function of $\delta$  are reported in Fig. \ref{fig:1}.
The solid and the broken lines correspond  to the curves $m_1(\delta)$,  derived, as explained before,  
from the two  conditions among real parameters obtained by \eqref{eq:delta1} and \eqref{eq:delta2}, respectively.
The solutions  $(m_1, \delta)$ correspond to the intersections between the two lines.
Exploiting the constraints on the imaginary parameters given by the same \refs{eq:delta1}{eq:delta2} results in predictions for $\alpha$ and $\beta$ as well.
Summarizing,  the yet-unknown neutrino parameters $m_1$, $\delta$, $\alpha$ and $\beta$ are given, in our approach, by the following two possibilities 
\begin{equation}
  \label{final:par}
m_1 \simeq 2.5 {\times}  10^{-3} \, \mathrm{eV} \qquad \delta \simeq \pm 0.6 \qquad \alpha \simeq \mp 1.4 \qquad \beta \simeq \mp 0.9
\eeq
which correspond to the upper or lower sign of the three phases.
\begin{figure}[t!!]
\begin{center}
\vspace{1cm}
\includegraphics[width=8.5cm,height=6cm,angle=0]{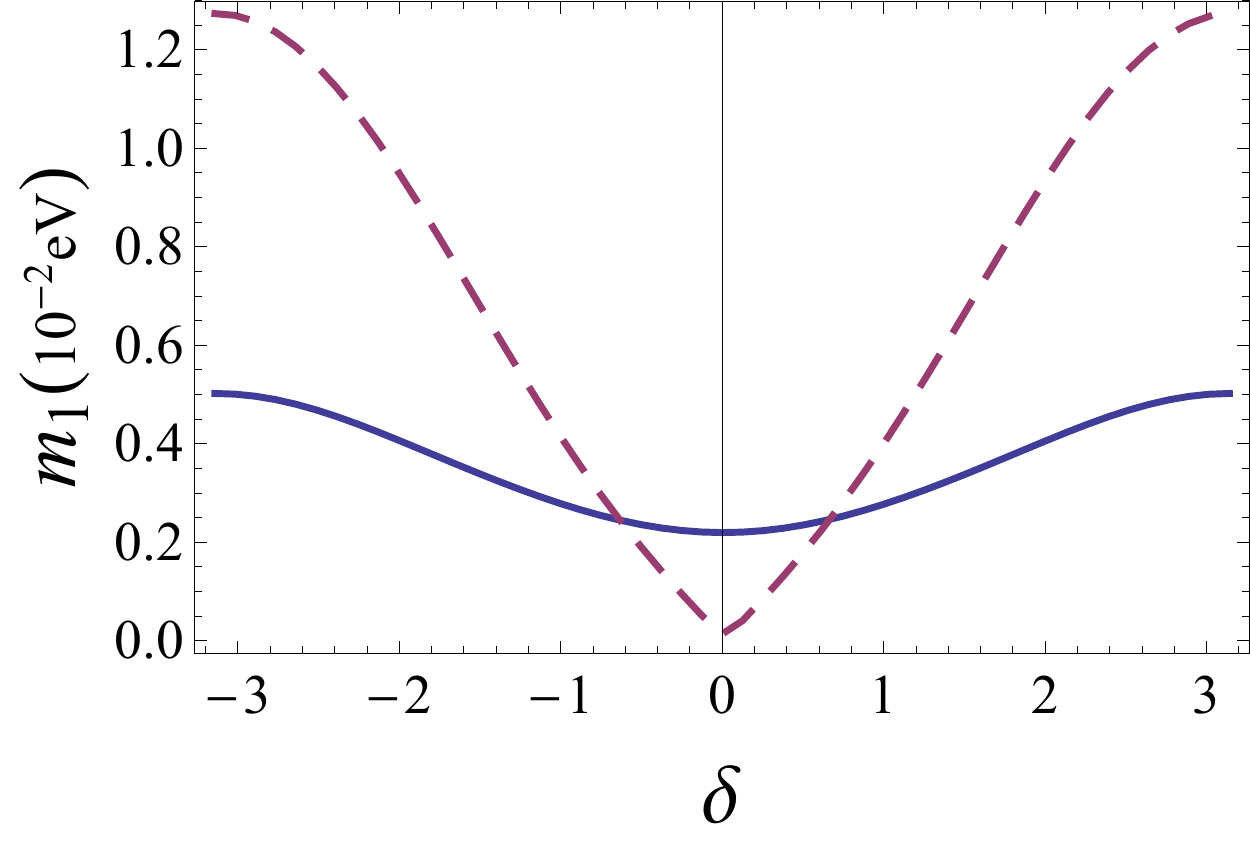}
\caption{Plots of $m_1$ in $meV$ as a function of $\delta$, when $(\theta_{12}, \theta_{23})=(-33.5^\circ, -42.3^\circ)$. The points of intersections represent possible solutions for $(m_1, \delta)$.
} \label{fig:1}
\end{center}
\end{figure}
Current experimental data have recently started to put constraints on the Dirac CP-violating
phase and we can compare with a recent result of  global 3$\nu$ oscillation analysis which give a 1$\sigma$ range $\delta/\pi \in [1.12, 1.77]$  for normal hierarchy \cite{Capozzi:2013csa}.
However, at $3\sigma$, all values $[0, 2]$ are still allowed.

\section{Leptogenesis}
\label{LEPTO}

Most of the interest in the values of the masses of RH neutrinos  lies in their double role in the see-saw mechanism and in leptogenesis. 
Without loss of generality, it is convenient to work in the basis
where the RH neutrino mass matrix $M_R$ is diagonal.  Since $M_{R}$ is
symmetric, it can be brought to diagonal form $M_{R}^{diag}={\rm
  diag}(M_1,M_2,M_3)$ with real and positive entries by means of a
unitary matrix $W$:
\begin{eqnarray}
M_{R}^{diag} & = & W^{\dagger}M_{R}W^{*}\,.
\label{eq:MR_d}
\end{eqnarray}
We indicate  the Dirac mass matrix 
in this basis as 
\begin{eqnarray}
\hat{m}_{D} & = & m_{D}W^{*} \label{eq:mdtilde}.
\end{eqnarray}
In this section we  discuss   the same case study  of Sect. \ref{LEO},  by setting $(\theta_{12}, \theta_{23})=(-|\theta_{12}|, -|\theta_{23}|)$.
By  arranging the
ordering of RH neutrino masses according to $M_1 < M_2 < M_3$, our predictions for the RH masses are
\beq
M_1 \simeq  3.5  {\times}  10^{9} \; \mathrm{GeV} \qquad M_2 \simeq M_3 \simeq  8.7 {\times}  10^{9} \; \mathrm{GeV} 
\eeq
The numerical differences between the absolute values of each pair of
solutions for $\delta$ are negligible. There is no large hierarchy 
between the masses, and the RH spectrum is compact, with values in the correct range for leptogenesis.
Let us observe that the degeneracy of the eigenstates $M_2 \simeq M_3$  is  lifted when the condition \eqref{eq:conditions} is only approximately satisfied.

The CP asymmetry in the decay of the RH neutrino $N_i$ ($i=1,2,3$)
to a lepton $\ell_\alpha$  ($\alpha=e,\mu,\tau$) is given by~\cite{Covi:1996wh, Buchmuller:1997yu,Anisimov:2005hr}
\begin{eqnarray}
\epsilon_{i\alpha} & = & \frac{1}{8\pi v^{2}}\sum_{k\neq i}
\frac{{\rm Im}\left[\left(\hat{m}_{D}^{\dagger}\right)_{i\alpha}
\left(\hat{m}_{D}\right)_{\alpha k}
\left(\hat{m}_{D}^{\dagger}\hat{m}_{D}\right)_{ik}\right]}
{\left(\hat{m}_{D}^{\dagger}\hat{m}_{D}\right)_{ii}}
f_{_{LV}}\left(\frac{M_{k}^{2}}{M_{i}^{2}}\right)
\nonumber \\
& + & \frac{1}{8\pi v^{2}}\sum_{k\neq i}
\frac{{\rm Im}\left[\left(\hat{m}_{D}^{\dagger}\right)_{i\alpha}
\left(\hat{m}_{D}\right)_{\alpha k}\left(\hat{m}_{D}^{\dagger}
\hat{m}_{D}\right)_{ki}\right]}
{\left(\hat{m}_{D}^{\dagger}\hat{m}_{D}\right)_{ii}}
f_{_{LC}}\left(\frac{M_{k}^{2}}{M_{i}^{2}}\right),
\label{eq:CP_asym}
\end{eqnarray}
where $v=174$ GeV is the EW VEV.  The loop functions are
\begin{eqnarray}
f_{_{LV}}(x) & = & \sqrt{x}\left[\frac{1-x}{\left(1-x\right)^{2}+
\left(\frac{\Gamma_{i}}{M_{i}}-x\frac{\Gamma_{k}}{M_{k}}\right)^{2}}+1-
\left(1+x\right)\log\frac{1+x}{x}\right],\nonumber \\
f_{_{LC}}(x) & = & \frac{1-x}{\left(1-x\right)^{2}+
\left(\frac{\Gamma_{i}}{M_{i}}-x\frac{\Gamma_{k}}{M_{k}}\right)^{2}},
\label{eq:f_and_g}
\end{eqnarray}
where \beq
\Gamma_{i} \equiv \frac{M_{i}}{8\pi v^{2}}
(\hat{m}_{D}^{\dagger}\hat{m}_{D})_{ii}\eeq
 is the total $N_i$ width.  
The first term in  eq.~\eqref{eq:CP_asym} comes from
 lepton-number-violating wave and vertex diagrams, while the second term
is from the  lepton-number-conserving (but lepton-flavour-violating) wave diagram.
 The rescaled decay width
\beq
{\widetilde m}_i \equiv
\frac{8\pi{ v^2}}{M_i^2}\Gamma{_i} =
\frac{(\hat{m}_D^\dagger\hat{m}_D)_{ii}}{  M_i},
\eeq
which is also known as the effective washout parameter,  
parameterizes conveniently the departure from thermal equilibrium of
$N_i$-related processes (the larger ${\widetilde m}_i$, the closer to
thermal equilibrium the decays and inverse decays of $N_i$ occur, thus
suppressing the final lepton asymmetry). 

\begin{table}[t!]
\begin{center}
\begin{tabular}{>{$}l<{$}  >{$}l<{$} >{$}l<{$} }
\hline \hline \\  [-6pt]
 &\mathrm{Washout}\;  \mathrm{projectors} &  \\  [+4pt]
\hline \hline \\ [-3pt]
P_{1e} \simeq 0.02 &  P_{1\mu}\simeq 0.42 & P_{1\tau} \simeq 0.56 \\ [+4pt]
P_{2e} \simeq 7.40 {\times}  10^{-5} &  P_{2\mu}\simeq 1.62 {\times}  10^{-3} & P_{2\tau} \simeq 0.99 \\ [+4pt]
P_{3e} \simeq 7.42 {\times}  10^{-5}  &  P_{3\mu}\simeq 0.42 & P_{3\tau} \simeq 0.99 \\ [+4pt]
\hline \hline \\  [-5pt]
 & \mathrm{Washout}\;  \mathrm{parameters} &  \\ [+4pt]
\hline \hline \\ [-3pt]
{\widetilde m}_1 \simeq 7.6 {\times}  10^{-2}\; \mathrm{eV} &
 {\widetilde m}_2 \simeq 565\; \mathrm{eV} &
 {\widetilde m}_3 \simeq 565\; \mathrm{eV} \\  [+4pt]
\hline
\end{tabular}
\caption{Leptogenesis washout projectors and parameters}
\label{tab:l1}
\end{center}
\vspace{-0.9cm}
\end{table}

The washout projector,
$P_{i\alpha}$,  projects the decay rate over the $\alpha$ flavour,
that is, it corresponds to the branching ratio for $N_{i}$ decaying to
$\ell_{\alpha}$, and can be written as
\begin{equation}
P_{i\alpha}=\frac{\left(\hat{m}_{D}^{\dagger}\right)_{i\alpha}
\left(\hat{m}_{D}\right)_{\alpha i}}
{\left(\hat{m}_{D}^{\dagger}\hat{m}_{D}\right)_{ii}}.
\label{eq:fla_proj}
\end{equation}
 Finally, the combination
$P_{i\alpha}\, {\widetilde m}_i$ projects the washout parameter over a
particular flavour direction, and determines how strongly the lepton
asymmetry of flavour $\alpha$ is washed out.

Our results for the washout projectors and  parameters are collected in Table \ref{tab:l1}, given the values found in  Eq. \eqref{final:par} (differences for $\delta >0$ or $\delta <0$ 
are negligible). Our
 results for the CP asymmetries  are collected in Table \ref{tab:l12},
for positive and negative values of $\delta$, respectively.

In order to calculate the baryon asymmetry, we need to solve a set of
Boltzmann equations (BE) derived as in Ref. \cite{Buccella:2012kc}. We report here such derivation for convenience's sake.
By including for simplicity only
decays and inverse decays, the BE for the RH neutrino densities
$Y_{N_i}$ and for $Y_{\Delta_\alpha}$, that is the asymmetry density
of the charge $B/3-L_\alpha$ normalized to the entropy density $s$,
take the form:
\begin{eqnarray}
sHz\frac{dY_{N_{i}}}{dz} & = & -\gamma_{N_{i}}
\left(\frac{Y_{N_{i}}}{Y_{N}^{eq}}-1\right),\nonumber \\
sHz\frac{dY_{\Delta_{\alpha}}}{dz} & = &
-\sum_{i}\left[\epsilon_{i\alpha}\gamma_{N_{i}}
\left(\frac{Y_{N_{i}}}{Y_{N}^{eq}}-1\right)
-\frac{\gamma_{N_{i\alpha}}}{2}
\left(\frac{Y_{\Delta\ell_{\alpha}}}{Y_{\ell}^{eq}}+
\frac{Y_{\Delta H}}{Y_{H}^{eq}}\right)\right],
\label{eq:BE}
\end{eqnarray}
where $Y_{N}^{eq}=\frac{45}{4\pi^{4}g_{*}}z^{2}{\cal K}_{2}(z)$ is the
equilibrium density for the RH neutrinos  with $g_*=106.75$ and ${\cal
  K}_{2}$ the second order modified Bessel function of the second
kind, $2Y_{\ell}^{eq}=Y_{H}^{eq}=\frac{15}{4\pi^{2}g_{*}}$ are
respectively the equilibrium densities for lepton doublets and for the
Higgs, and the integration variable is $z=M/T$ with $T$ the
temperature of the thermal bath.
  Here $Y_{\Delta_{\alpha}} \equiv Y_{\Delta
  B}/3-Y_{\Delta L_{\alpha}}$ where $Y_{\Delta L_{\alpha}}$ is the total
lepton density asymmetry in the $\alpha$ flavour which also includes
the asymmetries in the RH lepton singlets. 
\begin{table}[t!]
\begin{center}
\begin{tabular}{>{$}l<{$}  >{$}l<{$} >{$}l<{$} }
\hline \hline\\  [-6pt]
 &\mathrm{CP}\;  \mathrm{asymmetries} &  \\  [+4pt]
\hline \hline\\ [-3pt]
\epsilon_{1e} \simeq (-0.13, -0.03 ) {\times}  10^{-5} &
\epsilon_{1\mu} \simeq (-1.02,1.39 ) {\times}  10^{-5} &
\epsilon_{1\tau} \simeq (1.16, -1.37) {\times}  10^{-5}  \\ [+4pt]
\epsilon_{2e} \simeq (0.67, -1.01 ) {\times}  10^{-9} &
\epsilon_{2\mu} \simeq (1.77,-1.88 ) {\times}  10^{-8} &
\epsilon_{2\tau} \simeq (1.23, -1.31) {\times}  10^{-5} 
 \\ [+4pt]
\epsilon_{3e} \simeq (0.70, -1.02) {\times}  10^{-9} &
\epsilon_{3\mu} \simeq (1.85,-1.91) {\times}  10^{-8} &
\epsilon_{3\tau} \simeq (1.23, -1.31) {\times}  10^{-5}  \\ [+4pt]
\hline \\  [-5pt]
\end{tabular}
\caption{CP asymmetries, The first and second values in parenthesis refer to positive and negative values of $\delta$, respectively, as given by Eq. \eqref{final:par}.}
\label{tab:l12}
\end{center}
\vspace{-0.9cm}
\end{table}
  Since RH neutrinos  only
interact with lepton doublets, the right hand side  of the second equation of
eqs. \eqref{eq:BE} involves only the LH lepton doublets density
asymmetry in a given flavour $\alpha$,
$Y_{\Delta\ell_{\alpha}}=A_{\alpha\beta}Y_{\Delta_{\alpha}}$ with
$A_{\alpha\beta}$ the flavour mixing matrix~\cite{flavour0} given in Eq.
(\ref{eq:Amatrix}). In  equation \eqref{eq:BE} it is  also used
$Y_{\Delta H}=C_{\beta}Y_{\Delta_{\beta}}$ the Higgs density asymmetry
with $C_{\beta}$~\cite{spectator2}  given in (\ref{eq:Amatrix}) and
$\gamma_{N_{i\alpha}}=P_{i\alpha}\gamma_{N_{i}}$ (no sum over $i$).
The $A$ flavour mixing matrix and the $C$ vectors  in the relevant temperature regime are given
by~\cite{Nardi:2006fx}
\begin{eqnarray}
A & = & \frac{1}{2148}\left(\begin{array}{ccc}
-906 & 120 & 120\\
75 & -688 & 28\\
75 & 28 & -688\end{array}\right),\nonumber \\
C & = & -\frac{1}{358}\left(37,52,52\right).
\label{eq:Amatrix}
\end{eqnarray}
We have solved numerically the
BE in ~\eqn{eq:BE} and found the baryon asymmetry generated through leptogenesis according to the relation~\cite{Harvey:1990qw}
\begin{eqnarray}
Y_{\Delta B} & = & \frac{28}{79}\sum_{\alpha}Y_{\Delta_{\alpha}}\,.
\label{eq:Y_B}
\end{eqnarray}
Our average result is
\beq
Y_{\Delta B}\simeq  2.19 {\times}  10^{-10}
\eeq
which correspond to the input parameters 
in eq. \eqref{final:par} with positive $\delta$. By comparing with experimental data, we find it sufficiently close to the experimental value to be phenomenologically acceptable.
Indeed, recent combined 
Planck and WMAP CMB measurements \cite{Ade:2013zuv, Bennett:2012zja} yield, at 95\% c.l. 
\begin{equation}
\label{eq:YB_CMB}
Y_{\Delta B}^{P/WMAP}=(8.58 \pm 0.22) {\times}  10^{-11}.
\end{equation}
Let us underline that it is not a trivial result to recover the sign and the order of magnitude of the experimental data, given the high degree of predictability of our model.

Comparison with data allows us to discard the second possibility granted  by  \eqref{final:par}, corresponding to $\delta<0$,  which results in  a negative  value $Y_{\Delta B}\simeq -0.23 {\times}  10^{-11}$.
Let us observe that  a small difference of input parameters
can have a non negligible impact on the values of  leptogenesis  asymmetries, in contrast to what happens for the values of  masses $m_1$ and $M_i$. 

\section{Phenomenology in neutron-antineutron physics}

 The mass matrix $M_{RH}^{NP}$ has to have eigenvalues smaller than 
 the LR symmetry breaking scale $v_{R}$:
 $$M_{RH,1,2,3}^{E2}<v_{1,2}<v_{R}$$
 On the other hand, we have assumed that 
 $$M_{RH,1,2,3}^{E2} \gg \frac{v_{1}v_{2}}{M_{F2}}$$
 So, the scale $M_{F2}$ has to be
 $M_{F2} \gg 10^{9}\, \rm GeV$.
 This case is compatible with the natural situation $M_{F2}\simeq M_{S}$
 \footnote{ As a consequence, our model is not compatible with 
 a TeV-ish LR symmetric model}. 
 
 On the other hand, the string scale has necessary to be 
 higher than the RH neutrino  mass, {\it i.e}
 $M_{S}>10^{9}\, \rm GeV$.
 These bounds have important implications for other signatures 
 in phenomenology. 
 
Neutron-antineutron transitions
  generated by new physics at a scale $300\div 1000\, \rm TeV$
  can be tested in the next generation of experiments. In particular the AB-model predicts 
  this signature, even if the precise scale is unknown.
  The strength of neutron-antineutron transitions is 
\be{Gnnbar2}
G_{n{-}\bar{n}}\simeq \frac{g_{3}^{2}}{16\pi}\frac{f_{11}^{2}v_{2}}{M_{\Delta^{c}_{u^{c}u^{c}}}^{2}M_{\Delta^{c}_{d^{c}d^{c}}}^{2}M_{SUSY}\mathcal{M}'_{0}}
\ee
  where $f_{11}=\tilde{f}_{11}v_{1}/M_{2}$ with
$\tilde{f}_{11}$ Yukawa couplings $\tilde{f}_{11}v_{1}Q^{c}Q^{c} \Delta^{c}/M_{F2}$, including $f_{11}\Delta^{c}_{u^{c}u^{c}}u^{c}u^{c}$ and $f_{11}\Delta^{c}_{d^{c}d^{c}}d^{c}d^{c}$; $\Delta_{u^{c}u^{c}},\Delta_{d^{c}d^{c}}$
 are the sextets contained in $\Delta^{c}$. 
 This can be rewritten as the following bound on the sextets
$$\frac{1}{f_{11}^{2}}M_{\Delta_{u^{c}u^{c}}}^{2}M_{\Delta_{d^{c}d^{c}}}^{2}>\frac{(300\,TeV)^{5}v_{2}}{M_{SUSY}M_{S}e^{-S_{E2'}}}$$
A conservative assumption on the sextets, in order to avoid FCNCs bounds,
is 
$M_{\Delta_{u^{c}u^{c}}}\simeq M_{\Delta_{d^{c}d^{c}}}>100\, \rm TeV$
(with $f_{11}\simeq 1$). 
Calling $x=v_{2}/M_{SUSY}$, FCNCs bounds will constrains $M_{S},e^{-S_{E2'}},x$
  as
$$x^{-1}M_{S}e^{+S_{E2}'}>100\,\rm TeV$$   
at system with $M_{SUSY}>10^{9}\, \rm GeV$, $v_{1,2}<v_{R}$ and $M_{SUSY}\leq M_{S}$.
These bounds correspond to several different regions of the parameters space,
compatible with neutrino physics.  
As a consequence, our model provides a viable way to generate 
 a Majorana mass for the neutron testable in the next generation of experiments
 \footnote{Neutron-Antineutron transitions could be also an intriguing 
 test for new interactions, as discuss in \cite{Addazi:2015pia}.}.
  On the the other hand, the generation of such a $B-L$ violating operator can be dangerous 
in combination with $B+L$ violating sphalerons: they can wash-out an initial lepton number asymmetry generated by 
 RH neutrino  decays. Of course, they can regenerate the correct amount of baryon asymmetry through a post-sphaleron mechanism, 
 as discussed in \cite{ss4,BMt}.
 On the other hand, from a string theory prospective, it is reasonable to consider the case in which the strength of the effective operators coupling six quarks increases as a {\it dynamical field} from the early Universe to the present epoch. 
 Moduli stabilization is one of the most challenging problem in string theory, because it necessary involves non-perturbative effects such as fluxes and stringy instantons. In string theory, coupling constants, such $\alpha_{em}$ and so on,
 are functions of dynamical moduli $f(\phi_{i})$, that in turn have to be somehow stabilized. 
 However, in principle, moduli can undergo a slow cosmological evolution rather than being exactly constant in time. As a result, a slowly growing coupling can be naturally envisaged in string inspired models. 
 A natural {\it ansatz} can be a solitonic solution in time connecting to constant asymptotes.
 The naturalness of such a proposal is also supported by the fact that usually the dependence of coupling constants on moduli is of exponential type.
 In our case, we can suggest a solitonic solution growing from $G_{n\bar{n}}(t \ll t_{e.w}) \ll \bar{G}_{n\bar{n}}(t_{e.w} \ll \bar{t} \ll t_{BBN})$ 
 to $\bar{G}_{n\bar{n}}$, 
 where $\bar{G}_{n\bar{n}}$ is bounded by direct laboratory limits. 
 Under this general assumption, we also avoid cosmological limits from BBN (Big Bang Nucleosynthesis). 
 Let us remark that the moduli dependence of $G_{n\bar{n}}$ could enter from the non-perturbative mixing of 10-plets $\Delta$, 
 {\it i.e} in instantonic geometric moduli.  Of course, such a proposal deserves future investigations in global stringy models, beyond the purposes 
 of this paper.

\section{Conclusions and remarks}

In this paper, we have considered an
alternative see-saw mechanism 
produced by exotic instantons rather than 
by spontaneous symmetry breaking.
We have named this mechanism ``exotic see-saw'' mechanism, since exotic instantons generate the main contribution to the mass matrix of RH neutrinos.
We have embedded such a mechanism 
in an (un)oriented string model
with intersecting D-branes and E-branes, giving rise to a  
Pati-Salam like model in the low energy limit, 
plus extra non-perturbative couplings. 
The specific unoriented quiver theory 
that we have considered was largely inspired by the one suggested in \cite{Addazi:2015hka}.
The present model has a predictive power in low energy observables, not common
to other see-saw models. 

Our model 
makes precise predictions for low energy physics, 
 from the acquisition of $11$ inputs from neutrino physics. 
Seven  degrees of freedom
 parameterize the geometry of
the mixed disk amplitudes, {\it i.e} of $E2$-instanton intersecting 
$D6$-branes' stacks. 
We have reconstructed the seven geometric parameters associated to the exotic instanton
and we have
predictions to compare 
with the next generation of experiments.
This will allow to 
indirectly test if the $E2$-instanton considered really  dominates the mass terms in the neutrino sector.
We have considered a class of mixed disk amplitudes 
 producing a RH neutrino  mass matrix with quasi degenerate spectrum of eigenvalues. 
The compactness of the RH neutrino  spectrum is geometrically understood
in terms of mixed disk amplitudes and it is a favorable feature for predictability. 
As shown, this mechanism can also realize a successful baryogenesis 
through RH neutrino  decays. In our model, a $\theta_{13}\neq 0$ is 
compatible with leptogenesis and other neutrino physics bounds.
Our model is also suggesting other possible signatures in 
neutron-antineutron transitions \cite{Addazi:2015hka}.
On the other hand, our model is assuming a supersymmetry breaking scale
$M_{SUSY} \gg 1\, \rm TeV$ as well as a Left-Right symmetry scale $M_{LR} \gg 1\, \rm TeV$.
A possible discover of Supersymmetry of Left-Right symmetry at LHC or future high energy colliders
would rule out our model. 
In conclusion, our model provides a unifying picture 
of particles and interactions that will be indirectly tested 
from different low energy channels in neutrino physics, 
flavor changing neutral currents, neutron-antineutron transitions and LHC. 

\vspace{1cm} 

{\large \bf Acknowledgments} 

It is a pleasure to acknowledge interesting conversations with the participants in the XIV Marcel Grossmann Meeting in Roma (12-18 July 2015) during which this project was carried on. G.~R. thanks C.~ S. Fong for very useful and interesting discussions.
The work of A.~A. was supported in part by the MIUR research
grant ``Theoretical Astroparticle Physics" PRIN 2012CPPYP7.
The work of M.~B. was partly supported by the INFN network ``ST\&FI'' and by the {\it Uncovering Excellence} Grant ``STaI'' of the University of Rome ``Tor Vergata''.
The work of G.~R. was supported in part by MIUR under project  2010YJ2NYW
and INFN under specific initiative QNP.


\end{document}